\documentclass[]{aa}

\usepackage{graphicx}
\usepackage{float}
\usepackage{amsmath}
\bibpunct{(}{)}{;}{a}{}{,}

\begin{document}

\title{AVIATOR: Morphological object reconstruction in 3D\thanks{The AVIATOR code and a Jupyter notebook illustrating its use is publicly available at https://github.com/BirgitHa/AVIATOR.}}
\subtitle{An application to dense cores}
\titlerunning{AVIATOR}

\author{Birgit Hasenberger\inst{\ref{inst1}} 
\and Jo\~{a}o Alves \inst{\ref{inst1}, \ref{inst2}}}

\institute{Department for Astrophysics, University of Vienna, T{\"u}rkenschanzstra{\ss}e 17, 1180 Vienna, Austria \label{inst1} 
\and Radcliffe Institute for Advanced Study, Harvard University, 10 Garden Street, Cambridge, MA 02138, USA \label{inst2}}

\date{Received <date> /
Accepted <date>}

\abstract{Reconstructing 3D distributions from their 2D projections is a ubiquitous problem in various scientific fields, particularly so in observational astronomy. In this work, we present a new approach to solving this problem: a Vienna inverse-Abel-transform based object reconstruction algorithm AVIATOR. The reconstruction that it performs is based on the assumption that the distribution along the line of sight is similar to the distribution in the plane of projection, which requires a morphological analysis of the structures in the projected image. The output of the AVIATOR algorithm is an estimate of the 3D distribution in the form of a reconstruction volume that is calculated without the problematic requirements that commonly occur in other reconstruction methods such as symmetry in the plane of projection or modelling of radial profiles. We demonstrate the robustness of the technique to different geometries, density profiles, and noise by applying the AVIATOR algorithm to several model objects. In addition, the algorithm is applied to real data: We reconstruct the density and temperature distributions of two dense molecular cloud cores and find that they are in excellent agreement with profiles reported in the literature. The AVIATOR algorithm is thus capable of reconstructing 3D distributions of physical quantities consistently using an intuitive set of assumptions.}

\maketitle

\section{Introduction}
\label{sec:Intro}

\begin{figure}[t]
\begin{center}
  \resizebox{\hsize}{!}
  {\includegraphics{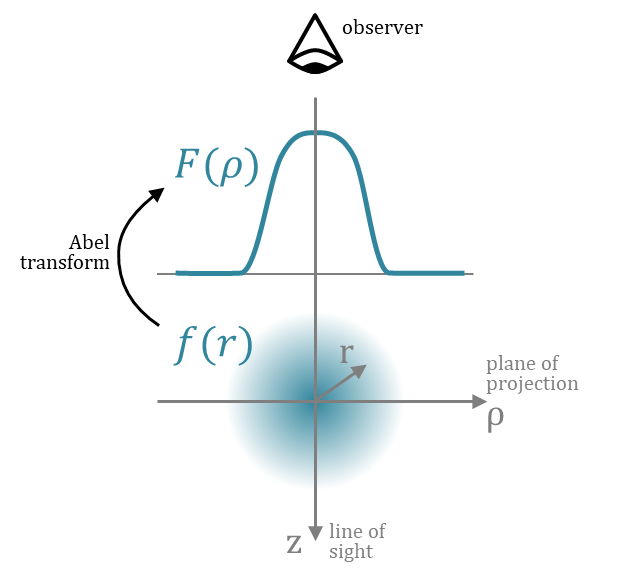}}
  \caption{Illustration of the Abel transform. For simplicity, the parameter $\rho$ represents the combination of the x- and y-axes, which together form the plane of projection.}
  \label{fig:Abelschematic}
\end{center}
\end{figure}

A fundamental aspect of many observational processes is the transformation of 3D to 2D information. For example, the observed 2D image of a 3D object with optically thin emission is obtained by integrating the emission along the line of sight. This type of transformation represents an inevitable part of many experimental measurement procedures. A commonly used mathematical characterisation of this conversion is the Abel transform \citep{Abel1826}, which describes the projection of a 3D axially symmetric distribution onto a 2D plane, as illustrated in Fig.~\ref{fig:Abelschematic}. In order to obtain physical 3D from observed 2D quantities, the inverse Abel transformation is used extensively in various fields, including studies on plasma spectra \citep[e.g.][]{Glasser1978}, photoionisation \citep[e.g.][]{Bordas1996}, the atmosphere of Earth \citep[e.g.][]{Kursinski1997} and other planets \citep[e.g.][]{Gladstone2016}, the graviational potential of galaxies \citep[e.g.][]{Binney1987}, and different phases of the interstellar medium \citep[e.g.][]{Bracco2017, Lee2015, Roy2014}.

In applications, the challenge lies in reconstructing a 3D object by deriving the inverse Abel transform from the observed 2D image. In its analytical form, the calculation of the inverse Abel transform requires assumptions on the 3D geometry of the object (e.g. spherical symmetry) and knowledge of the derivative of the radial 2D density profile. With real observational data, the presence of noise, estimating the accuracy of the geometric assumptions, extracting a radial density profile, and calculating its derivative can represent severe complications for the analysis \citep[e.g.][]{Minerbo1969, Craig1979, Roy2014}. To derive the inverse Abel transform while mitigating these complications, a number of numerical methods have been developed, some of which have recently been compared by \cite{Hickstein2019}.

In this work, we present a new method for reconstructing the 3D structure of an object from its 2D projection using the inverse Abel transform: AVIATOR, a Vienna inverse-Abel-transform based object reconstruction algorithm. The technique does not require any symmetry in the plane of the projection, nor the extraction of a density profile or the calculation of its derivative. Instead, the algorithm assumes that the morphology of the object along the line of sight is similar to its morphology in the plane of the projection and that it is mirror symmetric with respect to this plane. The AVIATOR algorithm allows for the reconstruction of objects using an intuitive set of assumptions, and ultimately, enables the consistent derivation of physical parameters in 3D.

One of many potential applications of this method are observations of molecular clouds and their internal structure. In particular, insights into the 3D morphology of dense cores, the birthplaces of stellar systems, can deepen our understanding of the star formation process. Three-dimensional properties of molecular clouds and dense cores derived from 2D observations were investigated by \cite{Steinacker2005}, \cite{Nielbock2012} \cite{Kainulainen2014}, \cite{Roy2014}, \cite{Krco2016}, \cite{Steinacker2016}, and \cite{Li2016}, for example, using vastly different techniques including multidimensional radiative-transfer modelling and Abel inversion. Here, we exemplify the use of the AVIATOR algorithm by applying it to dust emission data towards two molecular cloud cores and comparing the resulting 3D reconstruction with the literature. 

\section{Description of the algorithm}

Our approach uses the fact that the Abel inversion of a constant function can be expressed analytically. The strategy underlying the AVIATOR algorithm is therefore first to decompose the observed 2D image into structures of constant column density, second to apply the inverse Abel transform to each structure individually, and third to add the contributions of all structures to obtain the reconstructed 3D density distribution.

We assume a structure with constant column density of value $c$ up to a radius $R$. The column density distribution $F$ is then given as
\begin{equation*}
F(\rho) = c\, \theta(R-\rho),
\end{equation*}
where $\theta(x)$ is the Heaviside step function. When we use the inverse Abel transform and the assumption of spherical symmetry, the corresponding volume-density distribution $f$ can be derived as
\begin{equation}
f(r) = \frac{c}{\pi} \frac{1}{\sqrt{R^2-r^2}}.
\label{equ:iAt}
\end{equation}
However, the AVIATOR algorithm applies this inversion to each line of sight individually so that symmetry in the x-y plane is no longer required. Proof of the validity of this generalisation is given in Appendix~\ref{sec:proof}. In order to calculate the volume-density distribution, we thus require knowledge of three parameters at each point within a given structure: the column density, the maximum radius $R,$ and the radius $r$. Bypassing the assumption of spherical symmetry necessitates additional processing to extract values for the parameters $R$ and $r$. To this end, the AVIATOR algorithm uses information on the morphology of the column-density structure, which is described in detail in the following section.

\subsection{Definition and characterisation of structures}
\label{subsec:struct_char}

\begin{figure}[h]
\begin{center}
  \resizebox{\hsize}{!}
  {\includegraphics{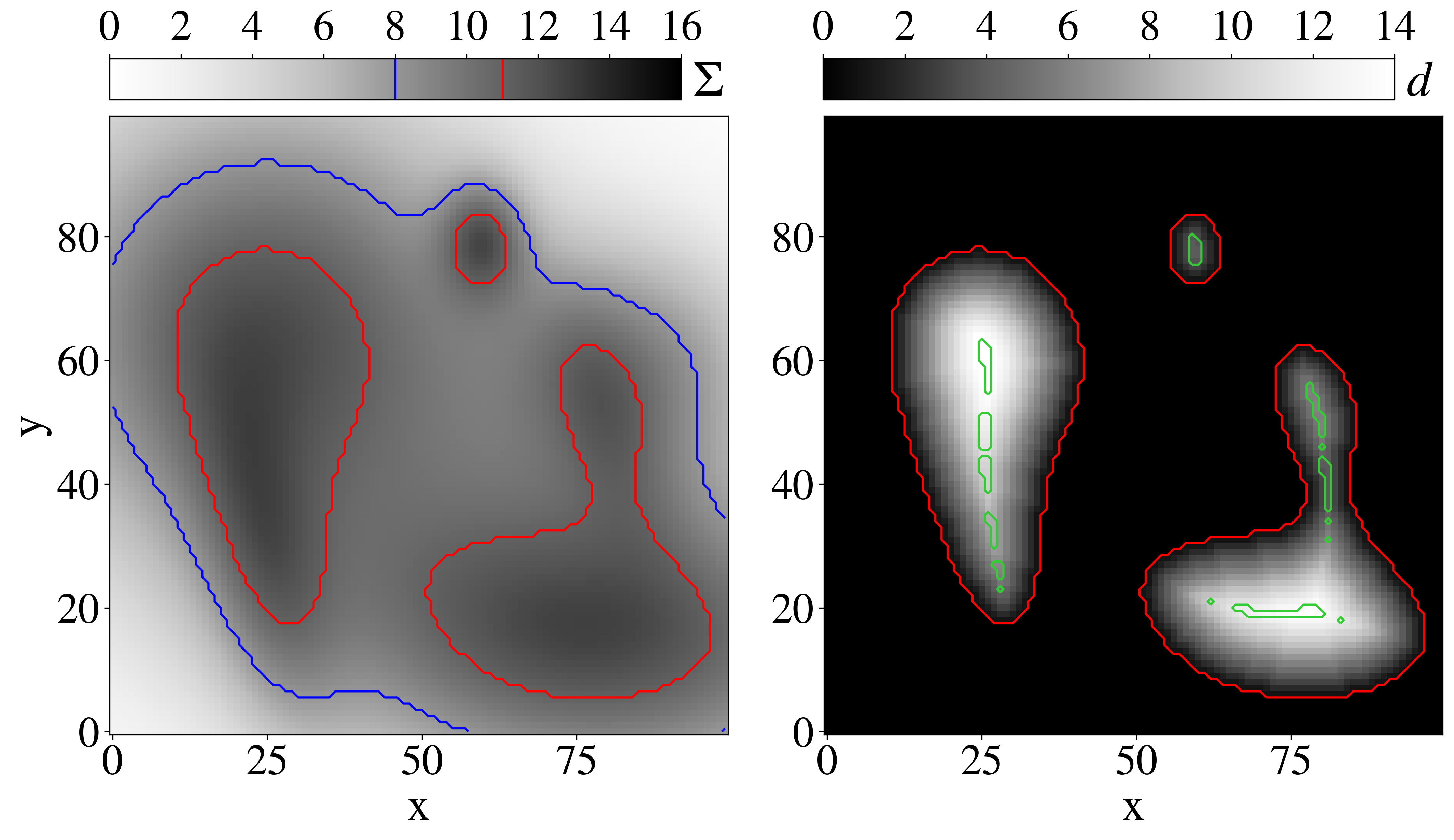}}
  \caption{\textit{Left:} Example column-density map consisting of several Gaussian distributions and two column-density levels highlighted as blue and red contours. \textit{Right:} Distance transform of the column-density level map corresponding to the red contour. Local maxima of the distance transform are indicated by green contours.}
  \label{fig:scheme_map_dist}
\end{center}
\end{figure}

To illustrate the morphological analysis of a column-density map as implemented in the AVIATOR algorithm, we constructed a simple column-density map consisting of several 2D Gaussian distributions (left panel in Fig.~\ref{fig:scheme_map_dist}) and show the result of key steps in the analysis. The goal of the procedure is to define structures and substructures in such a way that good estimates for the parameters $R$ and $r$ required in the inverse Abel transform can be found.

\begin{figure}[htb]
\begin{center}
  \resizebox{\hsize}{!}
  {\includegraphics{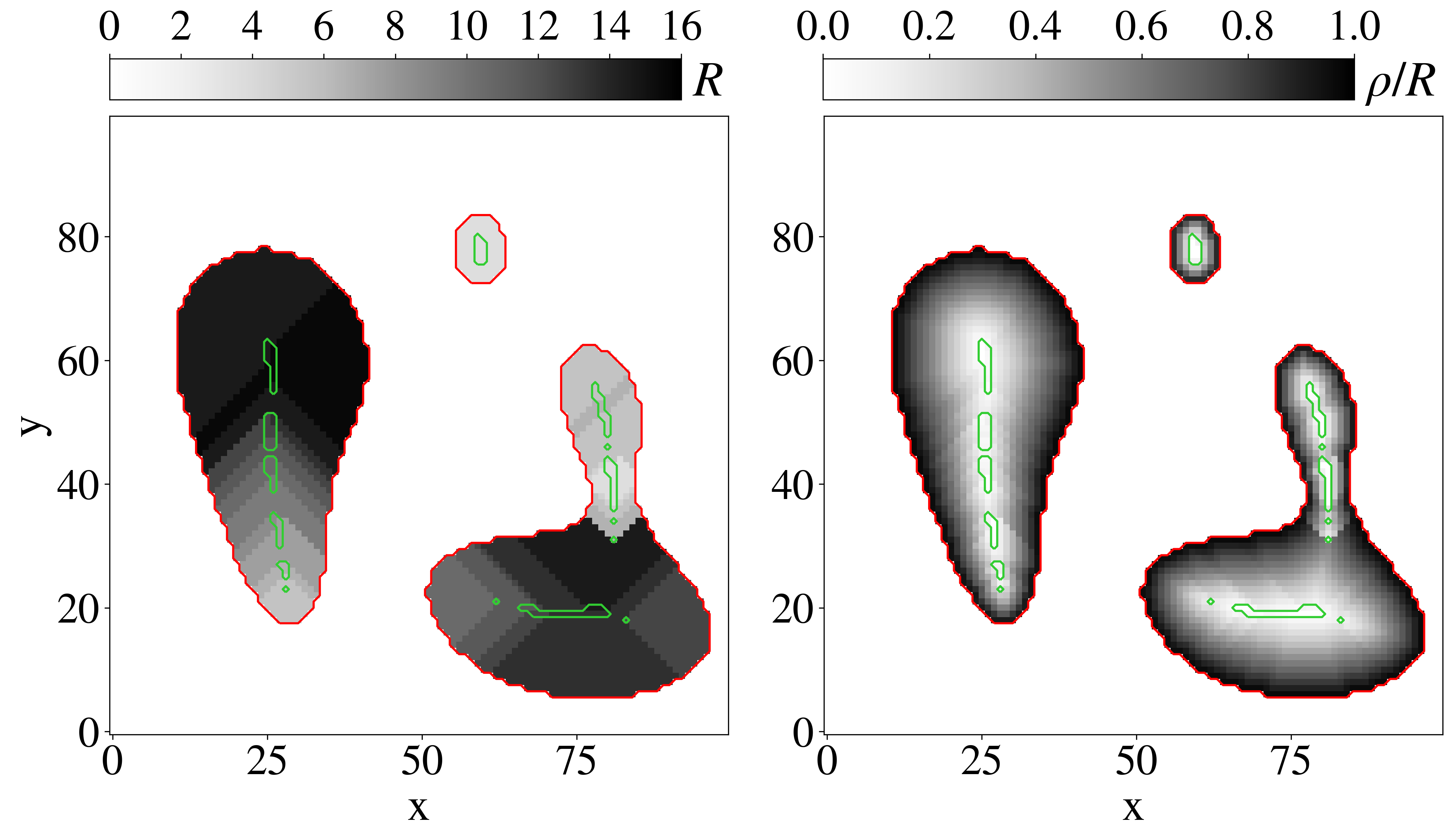}}
  \caption{\textit{Left:} Maximum radius $R$ for each substructure. \textit{Right:} Derived ratio $\rho/R$. Structures and local maxima of the distance transform are shown as in the right panel of Fig.~\ref{fig:scheme_map_dist}.}
  \label{fig:scheme_R_rat}
\end{center}
\end{figure}

First, the column-density map was decomposed into individual column-density levels (see left panel in Fig.~\ref{fig:scheme_map_dist}). This produced a series of maps, each containing structures of constant column-density, and the sum of this series of maps reproduced the original column-density map. Next, each level map was analysed individually. Because the structures in a level map might have complex shapes, we divided them into substructures using marker-controlled watershed segmentation. This algorithm separates a map into substructures in a similar way that a topographical map can be separated into catchment basins. Markers can be provided as a starting point for the segmentation process. We applied the watershed algorithm on the distance transform $d$ of the level map and defined the local maxima of this distance transform as input markers (see right panel in Fig.~\ref{fig:scheme_map_dist}). Each substructure can now be assigned a maximum radius $R$, which we define as the value of $d$ at the corresponding local maximum, and each pixel can be assigned a 2D radius $\rho$, which we define as $R$ subtracted by the distance transform (see Fig.~\ref{fig:scheme_R_rat}). The 3D radius $r$ for each voxel was derived via $r=\sqrt{\rho^2+z^2}$, where $z$ is the distance to the central plane of the reconstruction.

\subsection{Calculation of the inverse Abel transform for individual substructures}
\label{subsec:approx}

With the parameters $R$ and $r$ of the inverse Abel transform given in Eq.~\ref{equ:iAt} at hand, the next step was to calculate the volume-density distribution for each substructure. Because our calculations are based on a discrete grid of voxels with finite size, the expression in Eq.~\ref{equ:iAt} has to be integrated over the voxel volume. We derived an approximation for the volume-density distribution in Appendix~\ref{sec:derivations}. For voxels away from the centre of a substructure, 
\begin{equation}
f_{voxel}(x,y,z) \sim \frac{c}{\pi} \left[\arcsin\left(\frac{r_2}{R}\right) - \arcsin\left(\frac{r_1}{R}\right)\right],
\end{equation}
where $r_1$ and $r_2$ are the limits of the integration along the radial axis. They were set to the radius $r$ of the voxel subtracted by and added to half the length of a voxel side.

For the central voxels with $r=0$, the above approximation is not applicable. We therefore used a different relation, the derivation of which can be found in Appendix~\ref{sec:derivations}. For central voxels,
\begin{equation}
f_{voxel}^{central}(x,y,z) \sim \frac{c}{\pi R}(z_2-z_1) (y_2-y_1) (x_2-x_1),
\end{equation}
where each term in brackets corresponds to the difference of the integration limits and thus to the length of a voxel side. After we calculated the inverse Abel transform for each substructure, the contributions were summed to produce the final 3D reconstruction of the object.

\subsection{Assumptions and approximations}
\label{subsec:assump}

The starting point for the AVIATOR algorithm is the inverse Abel transform of a constant column-density profile for spherically symmetric structures as given in Eq.~\ref{equ:iAt}. The requirement of a constant profile is fulfilled if the original column-density map is decomposed into levels that are sufficiently close. In particular, a level map is guaranteed to contain structures with constant profiles if the list of decomposition levels is equivalent to the number of unique values in the column-density map. If this approach is not feasible because of memory or runtime constraints, fewer levels can be selected, which generally results in a loss of reconstruction quality.

Although the derivation of Equ.~\ref{equ:iAt} contains the assumption of spherical symmetry, we adapted the result in a way that allowed us to process other shapes as well. In Eq.~\ref{equ:iAt}, the parameter $r$ describes the distance to the central point of the object. For a substructure that is not the shape of a disc, a different definition of $r$ is required to ensure that the integrated volume-density distribution matches the original column-density map. To this end, we define $r$ via the distance transform, namely as the maximum radius $R$ subtracted by the distance to the edge of the structure (see Sect.~\ref{subsec:struct_char}). As a result of this definition, the resulting volume-density distribution of the substructure is not symmetric in the x-y plane, but only along the z-axis. Because $R$ is defined using the distance transform, the extent of a substructure along the z-axis is equivalent to the smallest extent of the substructure in the x-y plane. In the case of an ellipsoidal structure, this assumption is valid if the largest principal axis lies in the x-y plane. If this is not the case, the reconstructed volume-density distribution will over- or underestimate the real distribution depending on the inclination with respect to the x-y plane and the ratio of the principal axes (see also Sect.~\ref{subsec:ellipsoidal}).

Another assumption that is required to estimate the volume-density distribution is the position of substructures along the z axis.  In the current implementation, the AVIATOR algorithm assumes that the centres of all substructures are located in the same x-y plane, namely the central plane of the reconstruction. This behaviour might not always be desirable, for example if structures overlap in projection, but are known to be separated along the line of sight. In these cases, the contributions of the individual structures to the column-density map have to be separated and each structure reconstructed individually using the AVIATOR algorithm. The method aims at producing estimates of the volume-density distribution that are plausible based on the observed column-density distribution. The technique does not, however, attempt to separate structures along the line of sight because this information is not contained in column-density maps. We caution that any available information on the line-of-sight arrangement that is in conflict with the assumptions of the algorithm has to be considered before the algorithm is applied. To ensure a complete reconstruction of the column-density map, the total extent of the 3D reconstruction along the z-axis should be equal to the maximum radius $R$ of the largest structure in the map. For a structure that is centred in the column-density map, this radius corresponds to the sum of the map size in x and y.

\section{Validation of the reconstruction}
\label{sec:validation}

To evaluate the quality and robustness of the AVIATOR reconstruction, we performed tests on simulated objects with various shapes, sizes, density profiles, and noise levels. We modelled these objects in 3D, calculated their projected column-density distribution, applied the AVIATOR algorithm, and compared the reconstructed to the input distribution. In particular, we investigated the quality of the reconstruction regarding the overall 3D density distribution, the density profile, and the column-density map. All following tests were performed on a grid of $60\times60\times120$ voxels, and the volume-density distribution was normalised such that the peak value was one. The input and reconstructed radial volume-density profiles were extracted by defining shells with increasing inner radius and a thickness of one~pixel, and calculating the average volume density of voxels within these shells. The column-density maps for the input and reconstructed distributions were derived by summing the contributions along the z-axis.

We used three parameters to characterise the quality of the reconstruction: a) $f_{\Delta\rho/\rho_{in}}$, the fraction of voxels with a relative difference of $<10$\,\% between the modelled and reconstructed volume density, taking only voxels into account with $\rho_{in} > 0.01$ (equivalent to 1\,\% of the peak volume density), b) $\left( \Delta\Sigma/\Sigma_{in}  \right)_{max}$, the maximum relative deviation of the column-density maps, and c) $\left( \Delta\rho(r)/\rho_{in}(r) \right)_{max}$, the maximum relative deviation of the volume-density profiles. For all tests in this section, the derived values for these three parameters are listed in Appendix~\ref{sec:parametertables}.

The purpose of the following analyses is to demonstrate the reconstruction quality that can be achieved with the AVIATOR algorithm, quantified by a particular set of parameters, and for a particular set of models. These models were chosen to have comparably simple geometries and volume-density distributions. For many applications of the algorithm on observational data, these models will not be representative of the geometry and density distribution of the observed object. Therefore, the quality of reconstruction and possible systematic effects should be assessed individually for each application of the algorithm.

\subsection{Spherical density distributions}

\begin{figure}[tb]
\begin{center}
  \resizebox{\hsize}{!}
  {\includegraphics{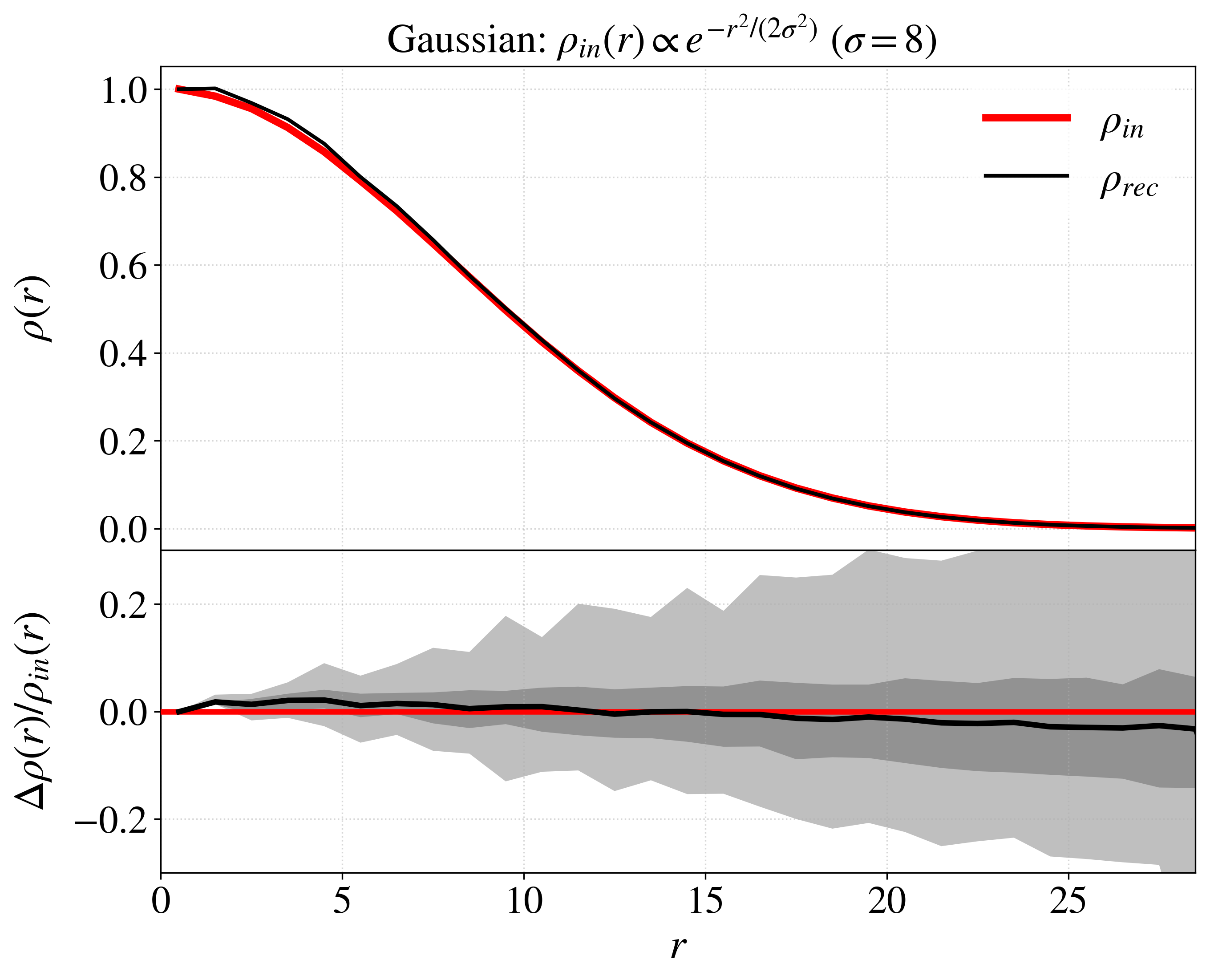}}
  \caption{Comparison of the input and reconstructed volume-density profile showing the radial profiles \textit{(upper panel)} and the relative difference \textit{(lower panel)} for a spherically symmetric density distribution. The light grey areas show the minimum and maximum reconstructed density value in each radial bin, and the dark grey areas show the first and third quartile.}
  \label{fig:val_sph_G8_prof}
\end{center}
\end{figure}

\begin{figure}[tb]
\begin{center}
  \resizebox{\hsize}{!}
  {\includegraphics{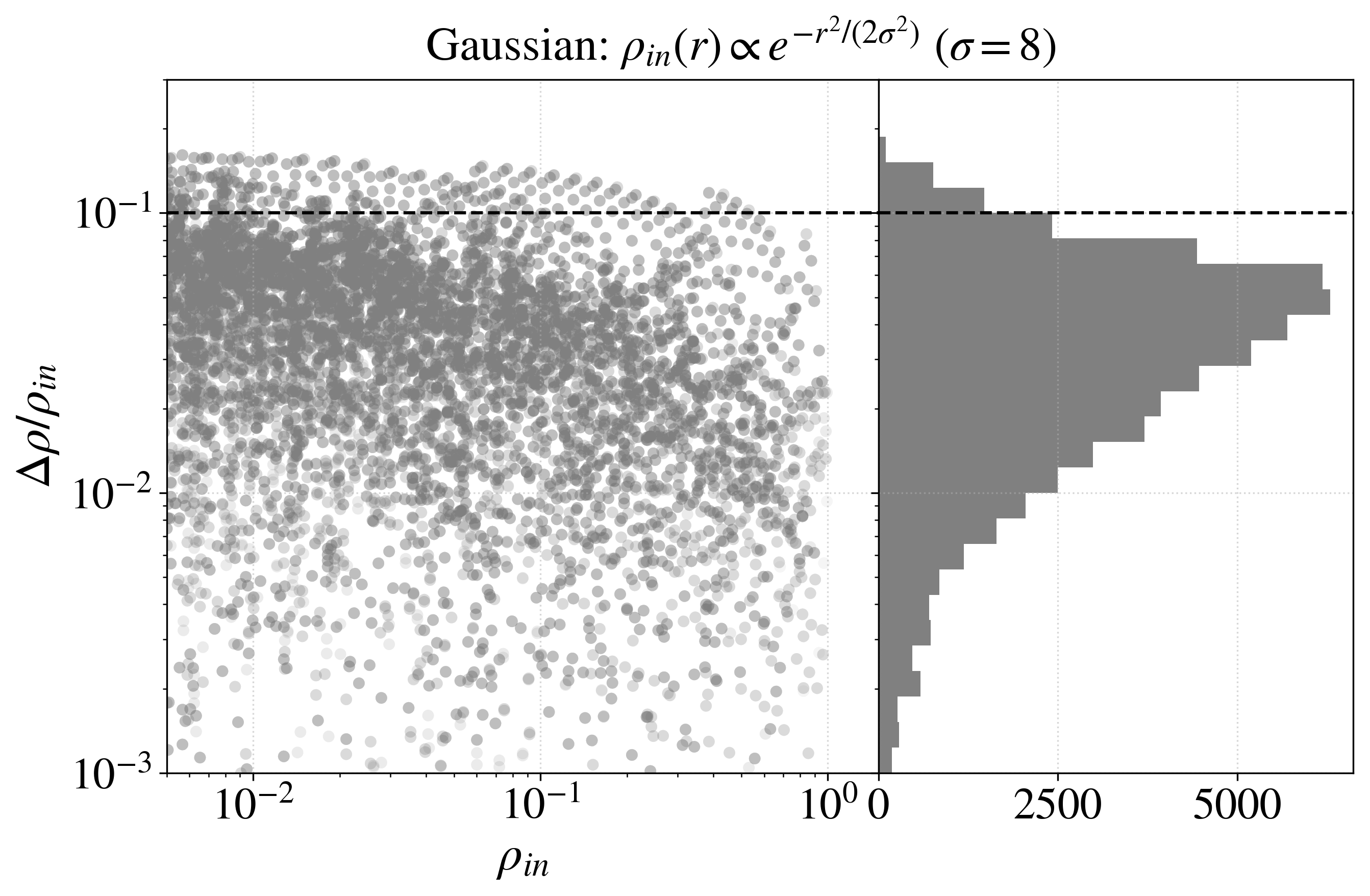}}
  \caption{Comparison of input and reconstructed volume densities showing the relative difference as a function of the input volume density \textit{(left)} and its histogram \textit{(right)} for a spherically symmetric density distribution. The dashed line corresponds to the limit that was chosen to define one of the reconstruction quality parameters, $f_{\Delta\rho/\rho_{in}}$, the fraction of voxels with small relative differences in volume density.}
  \label{fig:val_sph_G8_rho}
\end{center}
\end{figure}

\begin{figure}[tb]
\begin{center}
  \resizebox{\hsize}{!}
  {\includegraphics{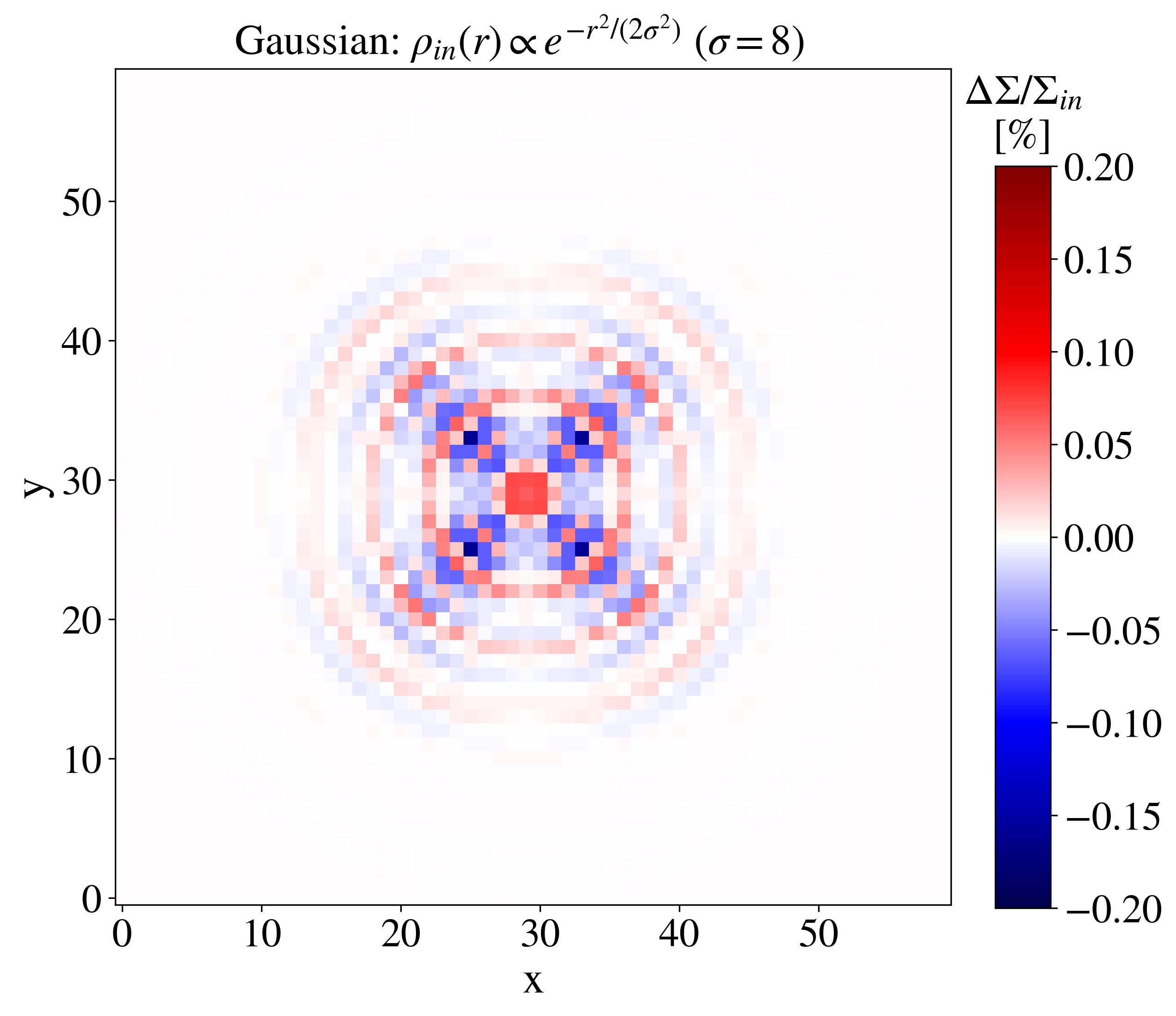}}
  \caption{Map of the relative difference between the input and reconstructed column-density map for a spherically symmetric density distribution.}
  \label{fig:val_sph_G8_Sigma}
\end{center}
\end{figure}

As a first test, we modelled objects with a spherically symmetric Gaussian density distribution. The standard deviation of the Gaussian function was varied between 3, 5, and 8~pixels. For all three values of the standard deviation, we find excellent agreement between the reconstructed and model density distributions (see Figs.~\ref{fig:val_sph_G8_prof} through~\ref{fig:val_sph_G8_Sigma} for graphs illustrating the reconstruction quality assuming $\sigma=8$ and Table~\ref{tab:sph_G} for a list of reconstruction quality parameters). The reconstructed volume-density profile and column-density map deviate by less than 7\,\% and 0.2\,\%, respectively, from the input distributions for all probed radii. The fraction of voxels with small relative deviations in their volume-density values is $>71$\,\%.  Generally, this fraction of voxels increases with higher standard deviation values of the Gaussian density profile because the transition region is resolved more accurately.

We repeated these tests with two other density distributions: a Plummer and a smooth step profile. Again, the parameters of the functions were varied ($R_P=5, 10, \text{and }15$~pixels for the Plummer profile and $a=0.5, 1.5,\text{and } 2.5$~pixels for the smooth step function) to modify the width of the transition. The reconstruction quality is similar to the Gaussian profile, as shown in Tables~\ref{tab:sph_P} and~\ref{tab:sph_s} of the appendix. Consequently, all following tests only include Gaussian density profiles.

\subsection{Ellipsoidal density distributions}
\label{subsec:ellipsoidal}

\begin{figure}[tb]
\begin{center}
  \resizebox{\hsize}{!}
  {\includegraphics{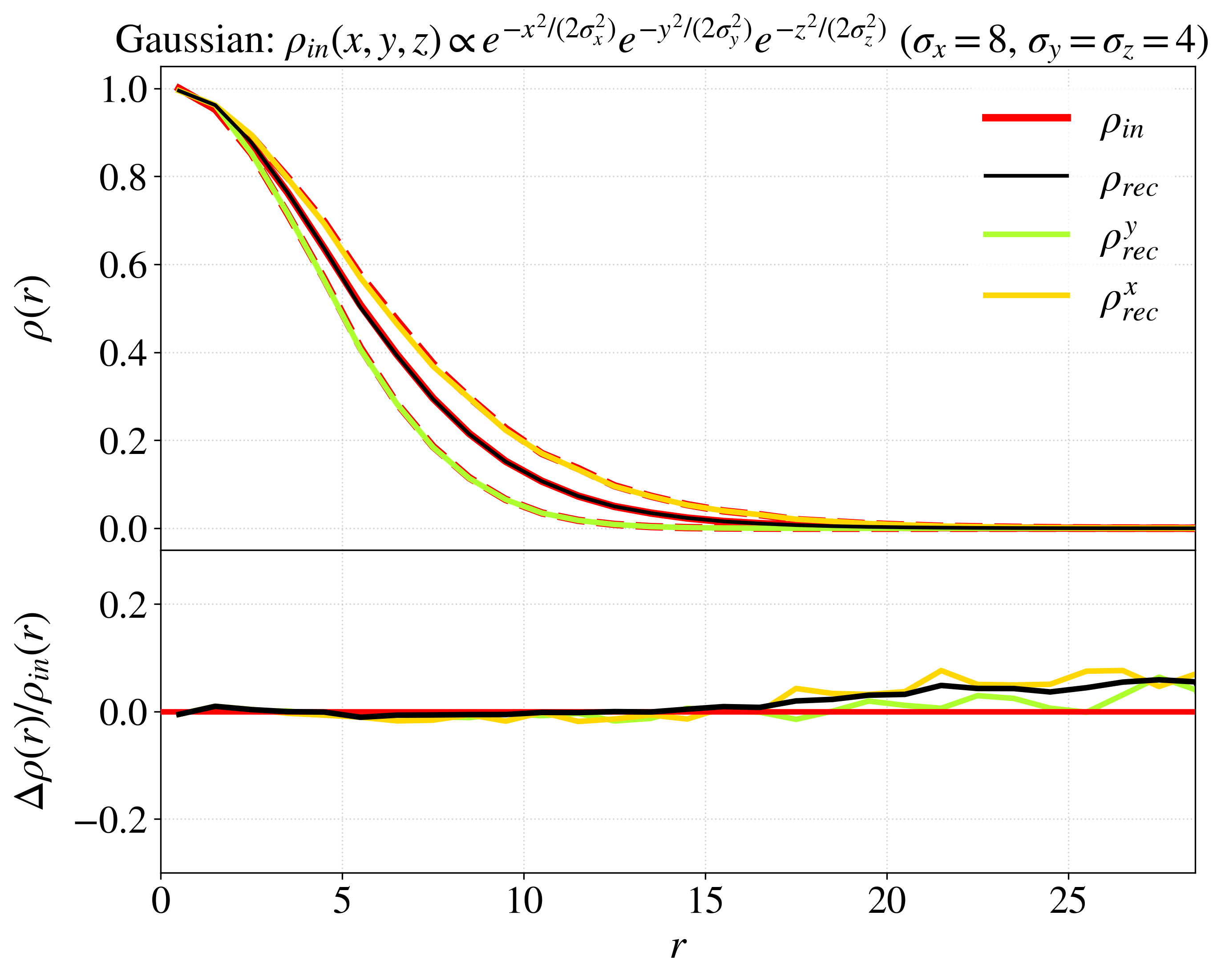}}
  \caption{Comparison of the input and reconstructed volume-density profile showing the radial profiles \textit{(upper panel)} and the relative difference \textit{(lower panel)} for a prolate density distribution. The yellow and green line correspond to profile extractions along the major and minor principal axis, respectively.}
  \label{fig:val_ell_G8_prof}
\end{center}
\end{figure}

\begin{figure}[tb]
\begin{center}
  \resizebox{\hsize}{!}
  {\includegraphics{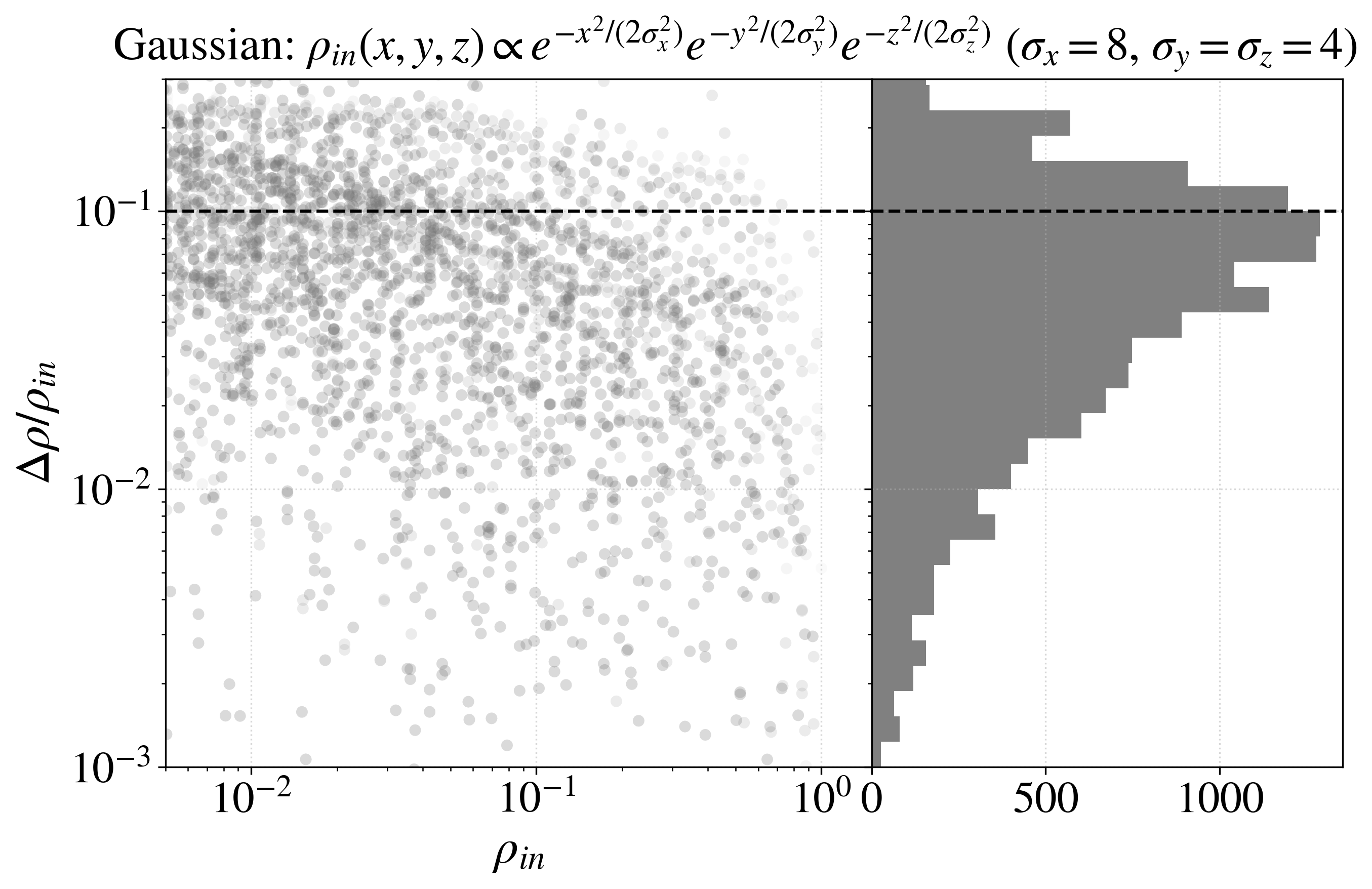}}
  \caption{Same as Fig.~\ref{fig:val_sph_G8_rho} for a prolate density distribution.}
  \label{fig:val_ell_G8_rho}
\end{center}
\end{figure}

\begin{figure}[tb]
\begin{center}
  \resizebox{\hsize}{!}
  {\includegraphics{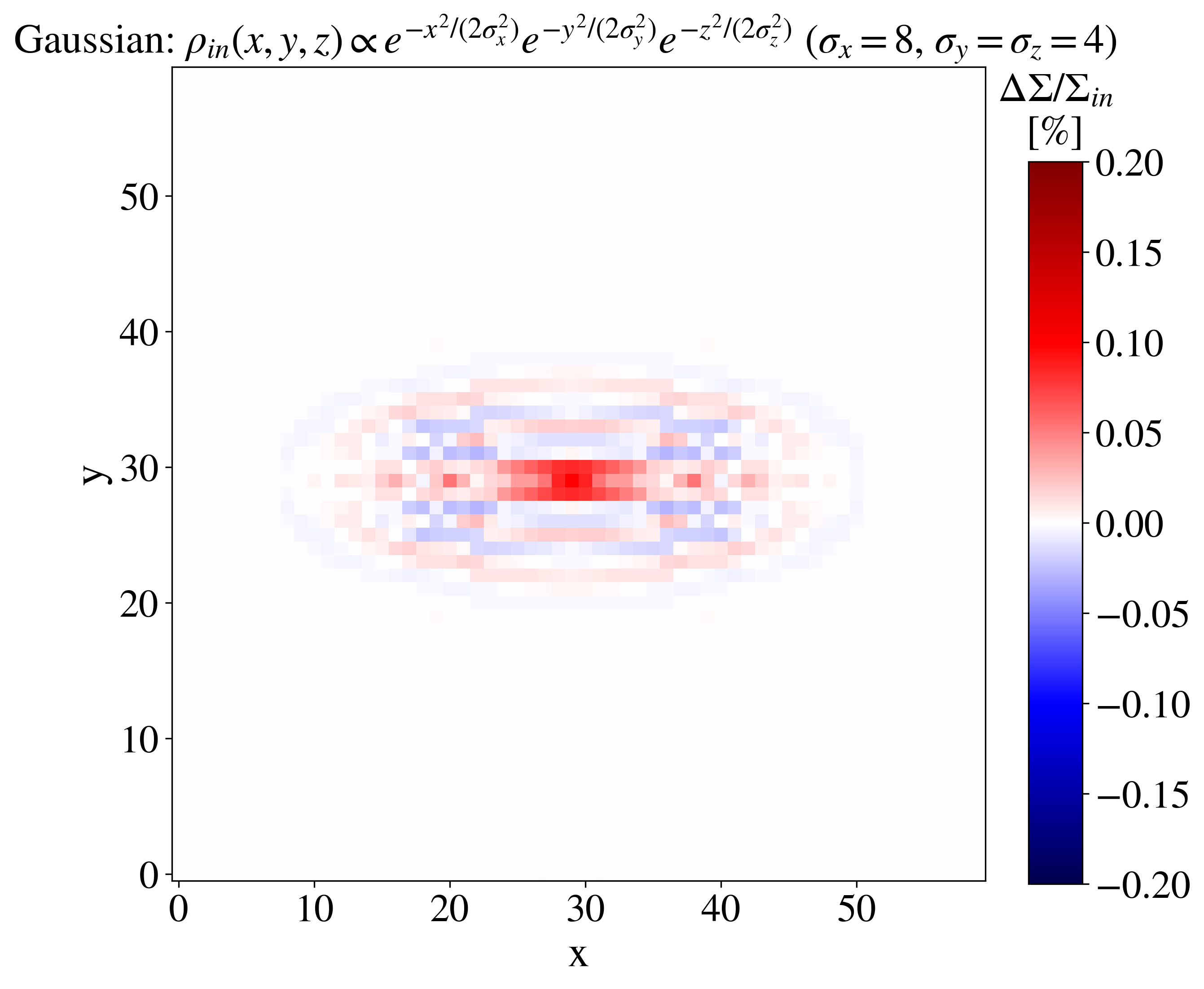}}
  \caption{Same as Fig.~\ref{fig:val_sph_G8_Sigma} for a prolate density distribution.}
  \label{fig:val_ell_G8_Sigma}
\end{center}
\end{figure}

Next, we considered objects with ellipsoidal density distributions. We first modelled a prolate object with the x-axis as its symmetry axis. In projection, this object appears as an ellipse with the semi-major and semi-minor axis along the x- and y-axes, respectively. We tested the quality of the reconstruction for $\sigma_x=8$ and three different values of the aspect ratio: $\sigma_x/\sigma_y=1.5, 2, \text{and }2.5$. In addition to the radially averaged density profile, we also extracted and compared the density distributions along the major and minor principal axes. Similar to the spherically symmetric case, the deviations in the density distribution as a whole, the deviations in the the density profiles, and those in the column-density map are small (see Figs.~\ref{fig:val_ell_G8_prof} through ~\ref{fig:val_ell_G8_Sigma} for graphs illustrating the reconstruction quality assuming $\sigma_x/\sigma_y=2$ and Table~\ref{tab:ell_G} for a list of reconstruction quality parameters). We find $f_{\Delta\rho/\rho_{in}} > 73$\,\%, $\left( \Delta\Sigma/\Sigma_{in}  \right)_{max} < 0.12$\,\%, and $\left( \Delta\rho(r)/\rho_{in}(r) \right)_{max} < 11$\,\%. The reconstruction quality per voxel exhibits the same trend as in the spherically symmetric case: The reconstruction is increasingly accurate for smoother transitions.

The spherical and spheroidal model objects described in the previous paragraphs fulfil the assumptions that the AVIATOR algorithm uses to reconstruct the density distribution because their extent along the z-axis is equivalent to the minimum extent in the x-y~plane. For ellipsoidal objects, this assumption is not generally fulfilled, for example, if it has an oblate shape, if the object is a tri-axial ellipsoid, or if it is inclined with respect to the x-y~plane. This information is lost during the projection of the 3D density distribution to the 2D column-density map, and the assumptions of the AVIATOR algorithm will lead to an over- or underestimation of the true volume density. The magnitude of this deviation is set by the difference between the true and assumed extent along the z-axis.

\begin{figure}[tb]
\begin{center}
  \resizebox{\hsize}{!}
  {\includegraphics{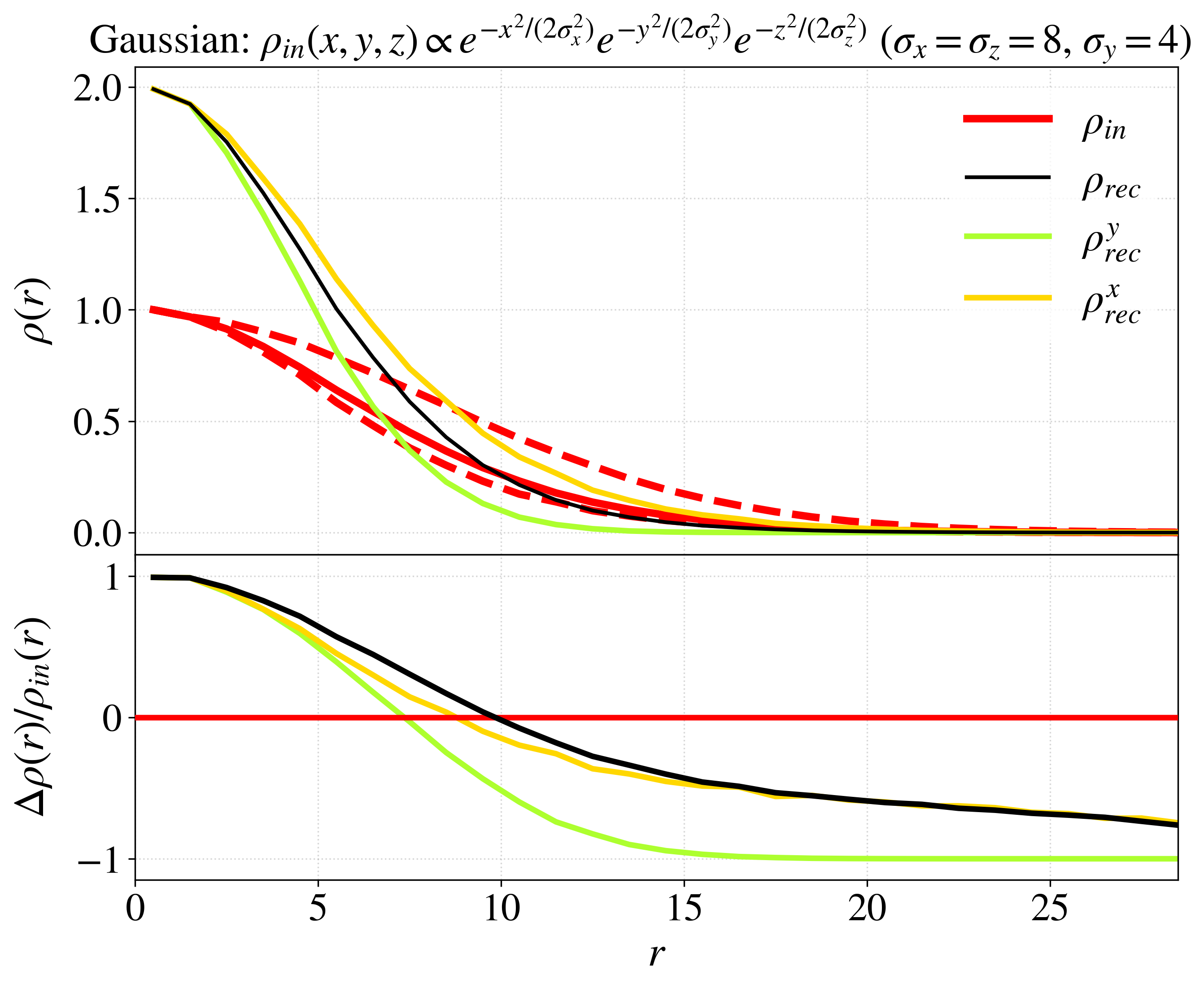}}
  \caption{Same as Fig.~\ref{fig:val_ell_G8_prof} for an oblate density distribution viewed edge-on.}
  \label{fig:val_ell_G8_oblate_prof}
\end{center}
\end{figure}

We tested this effect on an oblate object that is viewed edge-on, meaning that the minor principal axis lies in the x-y~plane and the extent along the z-axis is equivalent to the major principal axis. For simplicity, the minor principal axis was aligned with the y-axis in our model. We used the same values for the major principal axis and the range of aspect ratios as for the prolate case. As expected for these models, the AVIATOR algorithm overestimates the input volume densities in the central region of the object and underestimates them in the outer region (see Fig.~\ref{fig:val_ell_G8_oblate_prof}). At the very centre, the ratio between the reconstructed and the input volume density is equal to the aspect ratio. We can generalise this finding and state that the central density is over- or underestimated by a factor that is equal to the ratio of the true to the assumed extent along the z-axis.

\subsection{Column-density maps including noise}

Because observational data are always affected by noise, we also investigated the reconstruction quality for cases of noisy column-density maps. In its default setup, the AVIATOR algorithm analyses the morphology at every column-density level. Any noise will influence the shape of these column-density levels and therefore the reconstructed volume-density distribution. We tested whether this influence can be reduced when fewer levels are used to build the reconstruction.

To investigate the effect of noise and the choice of column-density levels on the reconstruction quality, we considered objects with a spherically symmetric Gaussian density distribution and a fixed standard deviation of 8~pixels. To the resulting column-density map, random noise sampled from a normal distribution was added, where the standard deviation $\sigma_{noise}$ was chosen as either 0.3 or 0.5. This corresponds to a mean signal-to-noise ratio in the column-density map of $\sim7.4$ and $\sim4.5$, respectively. We assessed the reconstruction quality by producing 100 different realisations of the noise in the column-density map per noise level and using the same metrics as in the previous analyses, averaged over the noise realisations. However, in this comparison we did not consider regions far from the object centre where the signal-to-noise ratio is low ($\rho_{in}/\sigma_{noise}<5$). An example of a noisy column-density map is shown in Fig.~\ref{fig:val_noise_G8_ex}. The reconstruction was performed with two different setups of the AVIATOR algorithm: a) the default setup that uses all column-density levels that exist in the map, and b) a setup with a custom list of column-density levels. We chose this list such that the minimum step size between column-density levels was equal to $2\sigma_{noise}$.

\begin{figure}[tb]
\begin{center}
  \resizebox{\hsize}{!}
  {\includegraphics{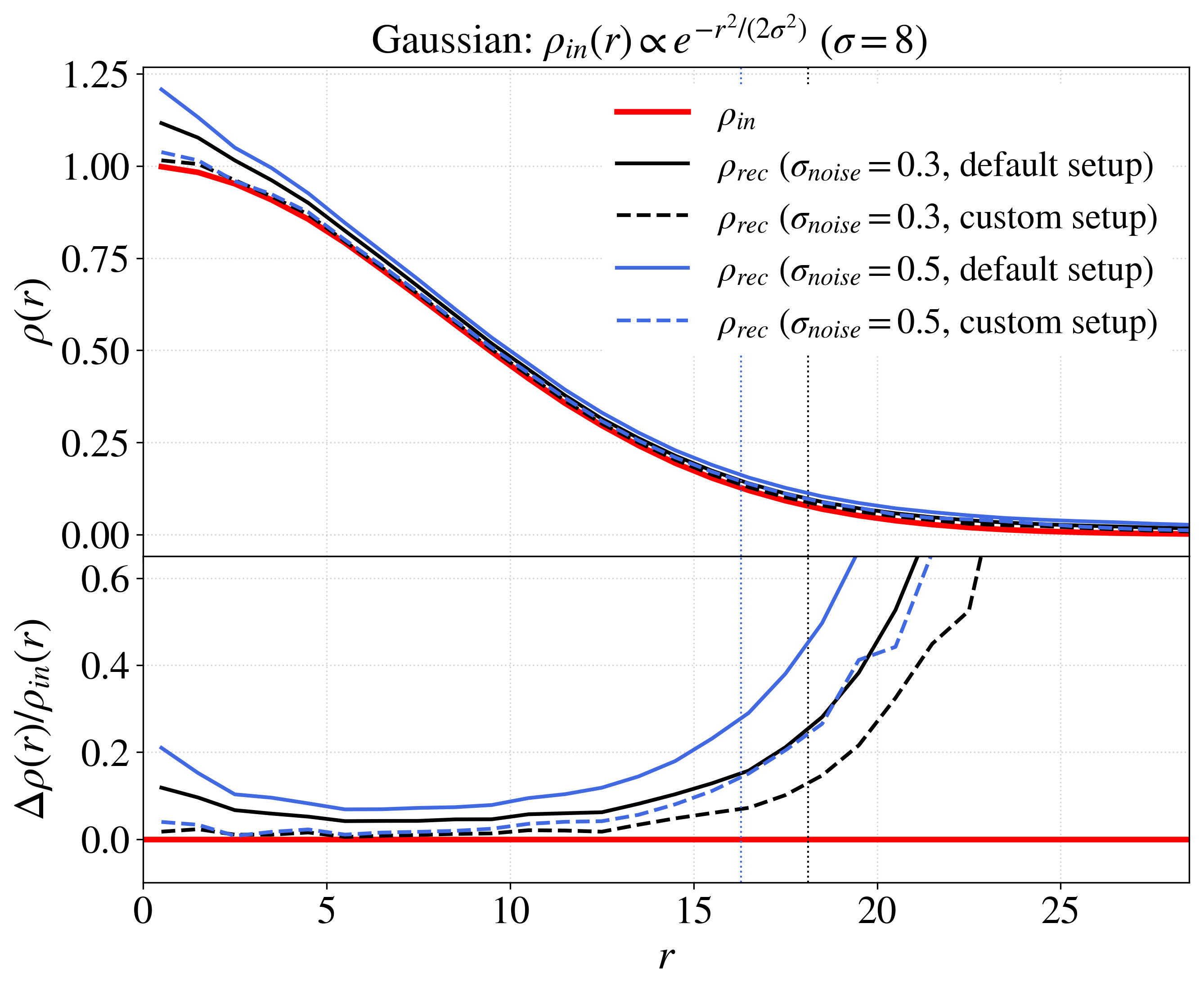}}
  \caption{Comparison of the input and averaged reconstructed volume-density profile showing the radial profiles \textit{(upper panel)} and the relative difference \textit{(lower panel)} for a spherical density distribution affected by different levels of noise in the column-density map. The vertical dotted lines indicate the radius at which the signal-to-noise ratio in the column-density map is 5.}
  \label{fig:val_noise_G8_prof}
\end{center}
\end{figure}

As expected, the models with a higher noise level generally exhibit a lower reconstruction quality (see Table~\ref{tab:noise_0p3} and~\ref{tab:noise_0p5} in the appendix). For example, using the default setup, $\left( \Delta\rho(r)/\rho_{in}(r) \right)_{max}=22$\,\% and $28$\,\% for $\sigma_{noise}=0.3$ and $0.5$, respectively, when averaged over the 100 realisations of the noise contribution. As shown in Fig.~\ref{fig:val_noise_G8_prof}, the reconstructed profiles overestimate the true profiles, which is a result of underestimating the extent of the structures in the column-density map. For both noise levels, the column-density map is reconstructed well, with deviations of $<0.5$\,\% on average.

When the custom setup of the AVIATOR algorithm is used, the reconstructed volume-density profile agrees with the input profile to a significantly higher degree. However, the reconstruction quality of the column-density map decreases and the fraction of voxels with small deviations from the true volume density decreases slightly. On the one hand, the reconstruction quality is increased in terms of the volume-density profile as a result of the reduced influence of noise on the reconstruction. On the other hand, the object morphology is traced on a coarser grid of column-density levels, which leads to a decrease in reconstruction quality regarding the column-density map and overall volume-density distribution. For this model, the most prominent improvement in its reconstructed profile can be observed in the central part of the object, which has a smaller effect on the quality parameters  $\left( \Delta\Sigma/\Sigma_{in}  \right)_{max}$ and $f_{\Delta\rho/\rho_{in}}$ because of the comparably small number of voxels it contains.

Overall, we find that noise in the column-density map has a negative effect on the reconstruction quality that can be alleviated to some extent by choosing an appropriate spacing of the column-density levels. The degree to which using a custom list of column-density levels improves the reconstruction of a noisy map depends on several factors, including the shape of the observed object, the choice of column-density levels, and the nature of the noise contribution. In this analysis, we do not consider any forms of noise reduction before applying the AVIATOR algorithm, for example using a smoothing filter on the column-density map. So many methods for reducing noise are available that a study that would test their influence on the AVIATOR reconstruction is beyond the scope of this paper.

\section{Application: dense molecular cloud cores}

One possible application for the AVIATOR algorithm is estimating 3D quantities from astronomical observations such as 2D flux maps. If the observed flux at any given point in the map is equivalent to the integrated flux along the line of sight and the geometry of the observed object is consistent with the assumptions described in Sect.~\ref{subsec:assump}, the AVIATOR algorithm can be used to estimate the 3D flux distribution.

To demonstrate this application, we used sub-millimetre observations of thermal dust emission towards molecular clouds. By combining data from the \textit{Herschel} and \textit{Planck} satellites, the $Herschel$-$Planck$-2MASS (HP2) survey provides images with high resolution and a high dynamic range of entire molecular cloud complexes: the Orion \citep{Lombardi2014}, Perseus \citep{Zari2016}, California \citep{Lada2017}, and Ophiuchus (Alves et al., \textit{in prep.}) molecular clouds, as well as the Pipe nebula \citep{Hasenberger2018}. For these regions, column-density and effective-temperature maps were derived through a modified-blackbody fit to the flux distribution on a pixel-by-pixel basis. By applying the AVIATOR algorithm and subsequently a modified-blackbody fit to portions of the corresponding flux maps, we can derive estimates of the 3D density and dust temperature distribution. We followed this scheme to extract radial density and temperature profiles for two dense molecular cloud cores, \object{B68} in the Pipe nebula and \object{L1689B} in the Ophiuchus molecular cloud complex. Profiles for these cores have been studied by \citet[][R14 in the following]{Roy2014} using the same observations and a more traditional application of the inverse Abel transform. These authors extracted radial flux distributions from the 2D maps, derived the inverse Abel transform numerically at each radius for various angular directions, and thus obtained density and temperature profiles. We compare these profiles to our AVIATOR results.

This first application of the AVIATOR algorithm to observational data was performed as follows: From the HP2 flux maps of the Pipe nebula and Ophiuchus complex, we extracted flux maps of B68 and L1689B, respectively, with a size of approximately 12$'$ and centred on the core. Next, we applied the AVIATOR algorithm to the flux maps in which the cores lie within the \textit{Herschel} coverage, using a custom list of levels to reduce the influence of noise. This produced a 3D flux distribution for each waveband. We then performed a modified-blackbody fit as described in the HP2 paper series to each voxel of the flux cubes, generating estimates of the 3D density and temperature distribution along with their respective errors. To ensure comparability with the results by R14, we assumed a uniform value of 2 for the dust emissivity index $\beta$ in the model. A column-density map was constructed by integrating the volume-density distribution along the z-axis, which represents the line of sight. The radial profiles were extracted in the same fashion as described in Sect.~\ref{sec:validation}. For this step and the calculation of the column-density map, we took only those voxels into account that exhibit small relative errors, that is, $\tau/\sigma_{\tau} > 3$ and $T/\sigma_T > 3$ for the density and temperature distributions, respectively. Here, the errors only represent the uncertainty due to the modified-blackbody fit. The centre of the core was set to the pixel with the highest value in the column-density map. To convert the column-density values derived by R14 into units of particles per $\mathrm{cm}^2$ and to account for differences in the assumed dust opacity law, we applied a conversion factor of 5.12 and 4.82 for B68 and L1689B, respectively. We estimated the uncertainty of the radial profiles by calculating the standard deviation of the reconstructed temperature or density within each radial bin. In the case of the R14 profiles, the errors correspond to the standard deviation of the reconstructed temperature or density for different angular directions. Thus, the given uncertainties do not include systematic effects.

\begin{figure}[tb]
  \resizebox{\hsize}{!}{\includegraphics{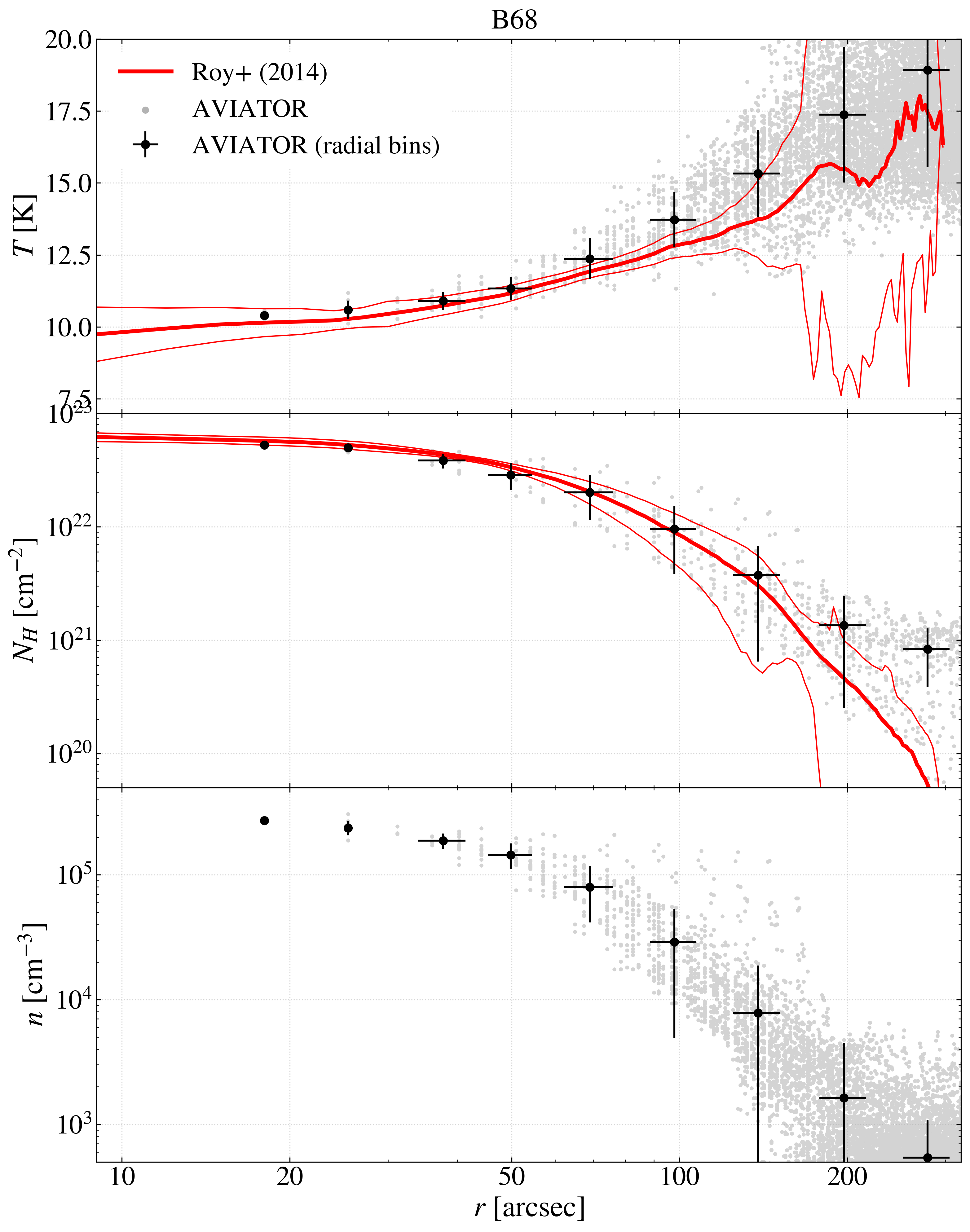}}
  \caption{Radial profiles in temperature \textit{(top)}, column density \textit{(centre)}, and volume density \textit{(bottom)} for the core B68. Grey circles show the results of the AVIATOR reconstruction. Black circles indicate the mean and standard deviation of the AVIATOR reconstruction per radial bin. The red lines show the mean and standard deviation of the reconstruction by R14.}
  \label{fig:B68profiles}
\end{figure}

\begin{figure}[tb]
  \resizebox{\hsize}{!}{\includegraphics{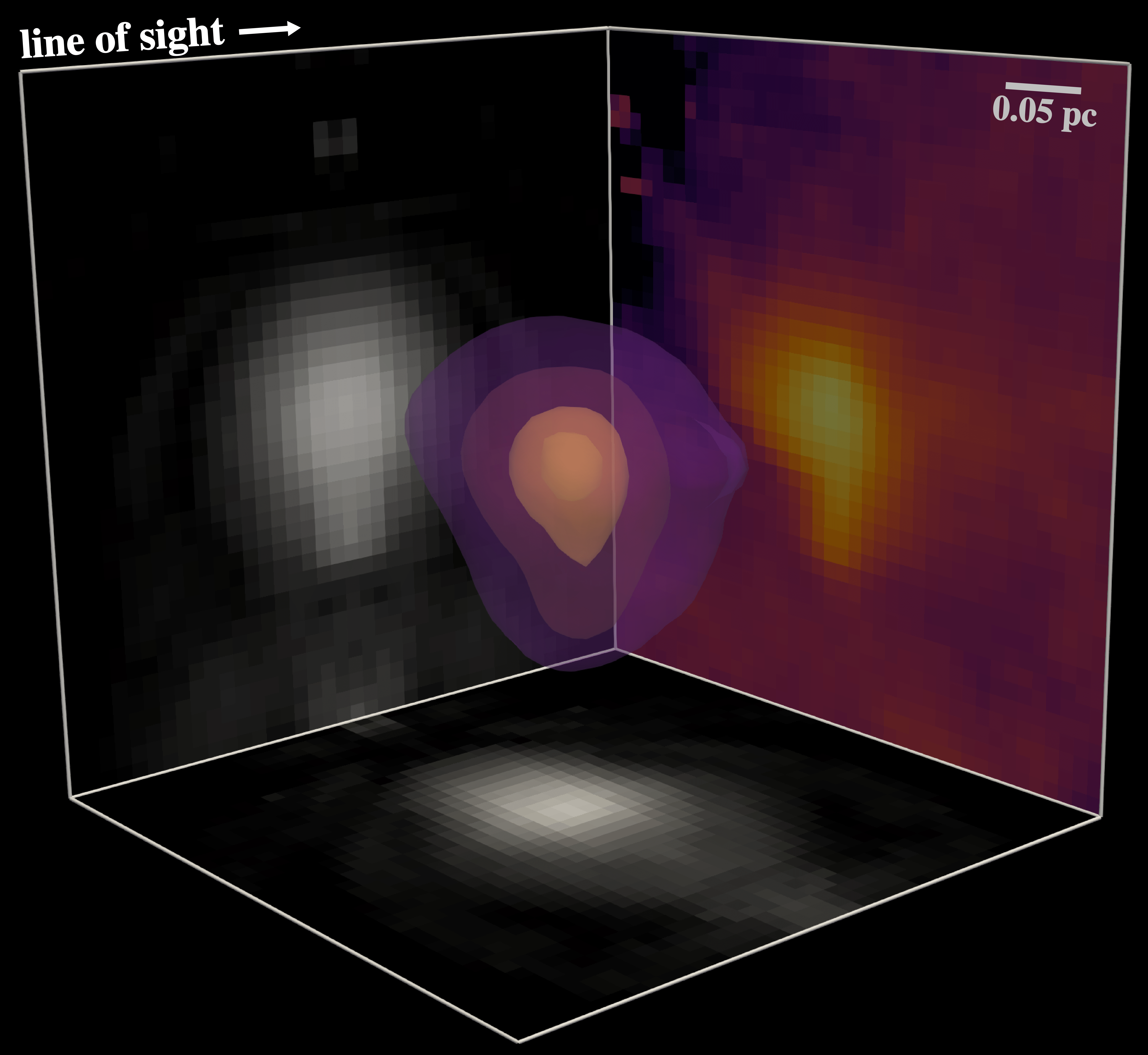}}
  \caption{Isocontours of the reconstructed volume-density distribution of the core B68, which is smoothed for visualisation puroses. The sides of the reconstruction cube show the projection along the axis perpendicular to that side. The projection shown in colour corresponds to the projection along the line of sight and thus exhibits a similar density distribution as observations of the core.}
  \label{fig:B683D}
\end{figure}

The comparison between the radial profiles as reported by R14 and those derived with the AVIATOR algorithm is shown in Fig.~\ref{fig:B68profiles} for the core B68. A 3D rendering of the reconstructed volume-density distribution is shown in  Fig.~\ref{fig:B683D}. The temperature and column-density distributions as reconstructed by the AVIATOR algorithm and their respective R14 counterpart are consistent within their errors up to a radius of $\sim200^{\prime\prime}$. However, the AVIATOR temperatures are systematically higher than the R14 temperatures. At the core centre (not shown in the figure), the maximum column-density and minimum temperature are derived as $(6.7\pm0.5)\cdot 10^{22}$\,cm$^{-2}$ and $9.3\pm0.5$\,K by R14, while we find $(6.0\pm0.3)\cdot 10^{22}$\,cm$^{-2}$ and $9.9\pm0.2$\,K. The maximum volume densities agree within their errors: they are $(3.8\pm0.3)\cdot 10^5$\,cm$^{-3}$ and $(3.7\pm0.3)\cdot 10^5$\,cm$^{-3}$ for the R14 and AVIATOR reconstruction, respectively. The uncertainties we report here for the central values of the AVIATOR profiles correspond to errors only due to the modified-blackbody fit.

\begin{figure}[tb]
  \resizebox{\hsize}{!}{\includegraphics{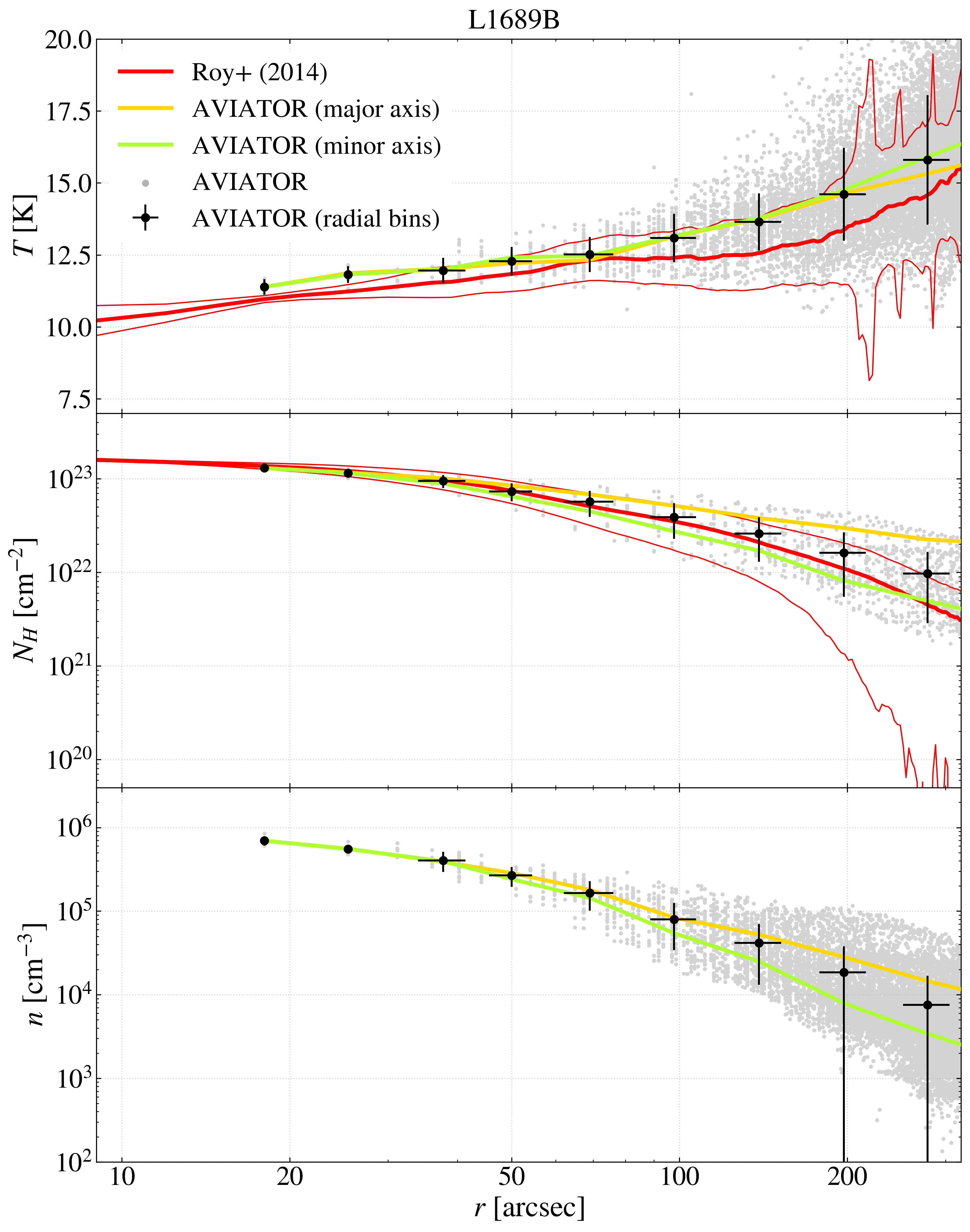}}
  \caption{Same as Fig.~\ref{fig:B68profiles} for the core L1689B. The yellow and green line correspond to profile extractions along the major and minor axis of the core, respectively.}
  \label{fig:L1689Bprofiles}
\end{figure}

\begin{figure}[ht]
  \resizebox{\hsize}{!}{\includegraphics{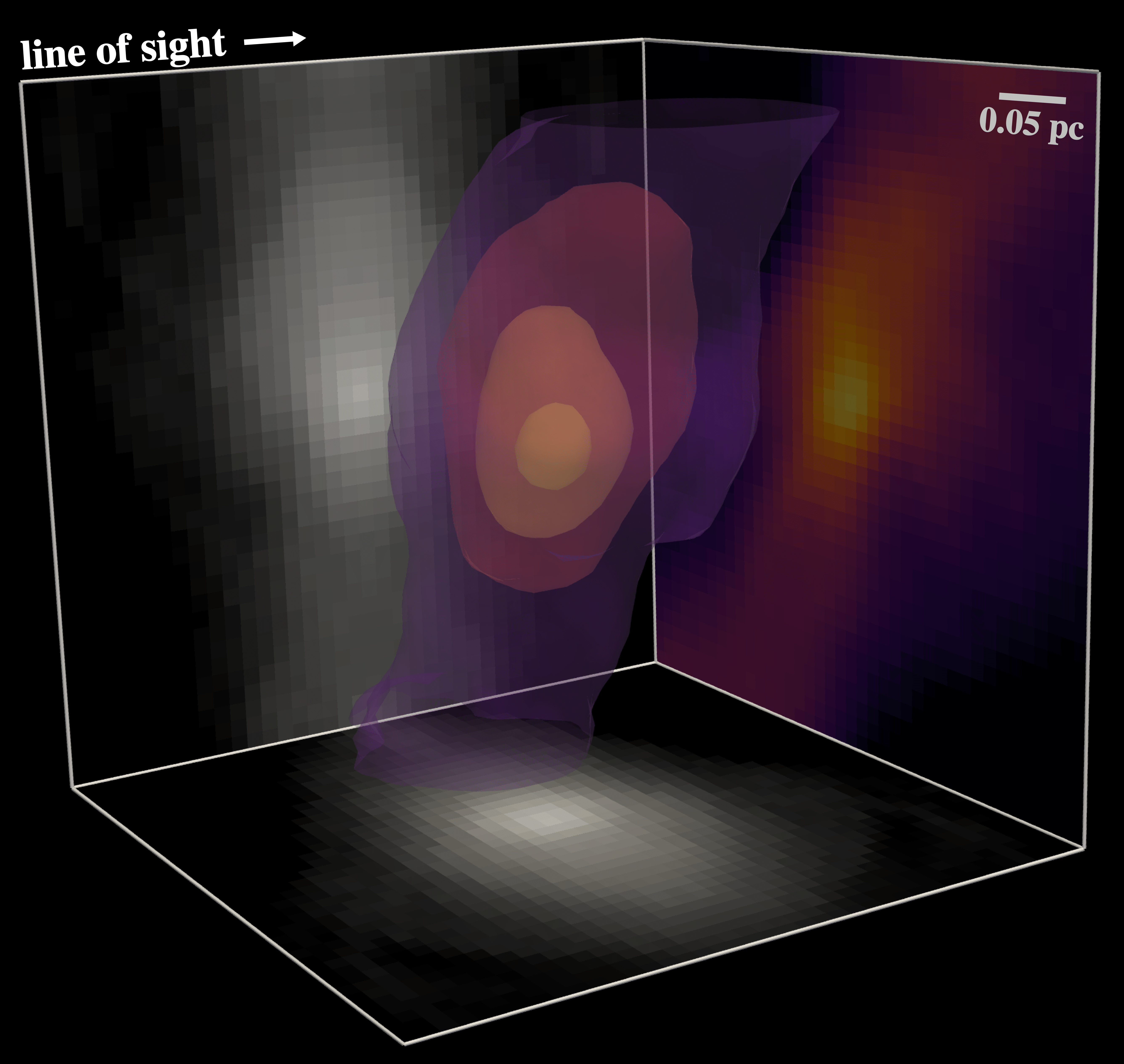}}
  \caption{Same as Fig.~\ref{fig:B683D} for the core L1689B.}
  \label{fig:L1689B3D}
\end{figure}

We compare the density and temperature profiles for the core L1689B in Fig.~\ref{fig:L1689Bprofiles} and present a 3D view of the volume-density distribution in Fig.~\ref{fig:L1689B3D}. The profiles are consistent within their errors away from the centre of the core, with the AVIATOR temperature profile lying systematically above the R14 profile. At the core centre, the maximum column and volume densities agree: R14 report values of $(9.5\pm0.5)\cdot10^{5}$\,cm$^{-3}$ and $(1.7\pm0.5)\cdot10^{23}$\,cm$^{-2}$, while we find $(9.5\pm1.0)\cdot10^{5}$\,cm$^{-3}$ and $(1.41\pm0.01)\cdot10^{23}$\,cm$^{-2}$. However, the minimum temperature is significantly higher for the AVIATOR reconstruction: it is $10.8\pm0.2$\,K compared to $9.8\pm0.5$\,K as derived by R14. Again, the uncertainties given for the AVIATOR reconstruction only represent the errors due to the modified-blackbody fit. We also show density and temperature profiles that were extracted along the filament in which the core is embedded (major axis of the core), and orthogonal to it (minor axis of the core). As expected for an elongated core, the volume and column-density profiles along these two directions are clearly different. The results are less clear for the temperature profiles, which appear similar over a wide range of radii. However, the temperature distribution exhibits a more complicated morphology that profiles along two directions might not be able to capture adequately, for example another less prominent temperature minimum towards the north (in Galactic coordinates) of the core centre.

Overall, we find a remarkable agreement between the temperature and density profiles of B68 and L1689B as reported by R14 and the AVIATOR reconstruction. Because the uncertainties given here do not consider any systematic effects, some of the discrepancies in the core centres could be negligible if these effects were taken into account. Systematic errors could be the result of various effects, including differences in the flux maps, for example different offset values, or in the reconstruction method itself, for example different assumptions regarding the line-of-sight geometry.

The density and temperature profiles of the cores B68 and L1689B have been investigated by \cite{Nielbock2012} and \cite{Steinacker2016}, respectively, using radiative transfer methods. While the central values for the temperature and for the column and volume densities agree within their errors for B68, these authors generally find lower temperatures and higher densities than we do in our analysis for the two cores. Radiative transfer modelling in 3D can produce estimates for density and temperature profiles based on a consistent treatment of dust emission, the incident radiation field (e.g. the interstellar radiation field and radiation from nearby massive stars), and its attenuation towards the core centre. In both studies, the \textit{Herschel} observations are complemented with additional data to expand the range of covered wavelengths. Deviations in the density and temperature profiles between these studies and the AVIATOR results could therefore be the result of the use of different datasets, data preparation techniques, and reconstruction methods. In the case of L1689B, for example, \cite{Steinacker2016} removed the contribution of the filament in which the core is embedded from the observed emission, which could lead to significant changes in the reconstructed parameters. Because 3D radiative transfer modelling and our approach differ in many aspects, a detailed investigation of the origin of the deviations in the resulting temperature and density is beyond the scope of this work.

Using the AVIATOR algorithm, we are able to reproduce density and temperature profiles extracted from \textit{Herschel} observations of dense cores using traditional Abel inversion methods. However, the AVIATOR algorithm provides a 3D estimate of density and temperature for a core in the form of a reconstruction volume rather than profiles along particular directions. This allows estimating physical quantities, such as the gravitational potential, in a way that takes the observed morphology of the core into account.

\section{Conclusions}

We presented a new method, the AVIATOR algorithm, for reconstructing 3D volume-density distributions from 2D column-density maps. The algorithm uses an intuitive set of assumptions. The technique models the volume density along the line of sight as similar to the distribution in the plane of projection by using a morphological analysis of each structure in levels of the column-density map. In contrast to many previous approaches reported in the literature that use the inverse Abel transform, the AVIATOR algorithm produces a model of the 3D density distribution in the form of a reconstruction volume and does not require symmetry in the plane of projection. Inherently, the technique depends on assumptions regarding the line-of-sight extent and arrangement of structures. To ensure a physically meaningful reconstruction, these assumptions must be applicable to the object at hand.

We tested the validity of our method on several simulated objects: The AVIATOR algorithm is capable of accurately reconstructing the volume-density distribution of spherically symmetric objects with a variety of radial volume-density profiles as well as prolate, ellipsoidal volume-density distributions with their major principal axis in the plane of projection. The reconstruction quality is reduced for objects that do not exhibit the assumed distribution along the line of sight. While noise in the column-density maps affects the reconstruction quality significantly, its effect can be alleviated by adapting the column-density level spacing. The AVIATOR algorithm was also applied to observations of dense molecular cloud cores to demonstrate its consistency with traditional methods that employ the inverse Abel transform. Using a similar data set, we find excellent agreement between the temperature and column-density profiles as derived in the literature and as found by applying the AVIATOR algorithm.

We conclude that this method is a robust and versatile tool for reconstructing 3D density distributions with many possible areas of application. The AVIATOR algorithm can therefore be a means to derive 3D physical parameters across a range of spatial scales in a more consistent and reliable fashion than was possible before.

\begin{acknowledgements}
We thank the anonymous referee for their comments which helped improve the paper. We would also like to thank J\"urgen Steinacker for his valuable feedback and Philippe Andr\'{e} for providing the column-density and temperature profiles presented by \cite{Roy2014}. This research made use of Astropy \citep{astropy2013}, numpy \citep{numpy2011}, scipy \citep{scipy2001}, skimage \citep{skimage2014}, matplotlib \citep{matplotlib2007}, and Paraview \citep{paraview2005}. 
\end{acknowledgements}

\bibliographystyle{aa}
\bibliography{mybib}

\begin{thebibliography}{28}
\expandafter\ifx\csname natexlab\endcsname\relax\def\natexlab#1{#1}\fi

\bibitem[{{Abel}(1826)}]{Abel1826}
{Abel}, N.~H. 1826, Journal f{\"u}r die reine und angewandte Mathematik, 1, 153

\bibitem[{Ahrens {et~al.}(2005)Ahrens, Geveci, \& Law}]{paraview2005}
Ahrens, J., Geveci, B., \& Law, C. 2005, in Visualization Handbook, ed. C.~D.
  Hansen \& C.~R. Johnson (Burlington: Butterworth-Heinemann), 717 -- 731

\bibitem[{{Astropy Collaboration} {et~al.}(2013){Astropy Collaboration},
  {Robitaille}, {Tollerud}, {Greenfield}, {Droettboom}, {Bray}, {Aldcroft},
  {Davis}, {Ginsburg}, {Price-Whelan}, {Kerzendorf}, {Conley}, {Crighton},
  {Barbary}, {Muna}, {Ferguson}, {Grollier}, {Parikh}, {Nair}, {Unther},
  {Deil}, {Woillez}, {Conseil}, {Kramer}, {Turner}, {Singer}, {Fox}, {Weaver},
  {Zabalza}, {Edwards}, {Azalee Bostroem}, {Burke}, {Casey}, {Crawford},
  {Dencheva}, {Ely}, {Jenness}, {Labrie}, {Lim}, {Pierfederici}, {Pontzen},
  {Ptak}, {Refsdal}, {Servillat}, \& {Streicher}}]{astropy2013}
{Astropy Collaboration}, {Robitaille}, T.~P., {Tollerud}, E.~J., {et~al.} 2013,
  \aap, 558, A33

\bibitem[{{Binney} \& {Tremaine}(1987)}]{Binney1987}
{Binney}, J. \& {Tremaine}, S. 1987, {Galactic dynamics}

\bibitem[{Bordas {et~al.}(1996)Bordas, Paulig, Helm, \& Huestis}]{Bordas1996}
Bordas, C., Paulig, F., Helm, H., \& Huestis, D.~L. 1996, Review of Scientific
  Instruments, 67, 2257

\bibitem[{{Bracco} {et~al.}(2017){Bracco}, {Palmeirim}, {Andr{\'e}}, {Adam},
  {Ade}, {Bacmann}, {Beelen}, {Beno{\^i}t}, {Bideaud}, {Billot}, {Bourrion},
  {Calvo}, {Catalano}, {Coiffard}, {Comis}, {D'Addabbo}, {D{\'e}sert},
  {Didelon}, {Doyle}, {Goupy}, {K{\"o}nyves}, {Kramer}, {Lagache}, {Leclercq},
  {Mac{\'{\i}}as-P{\'e}rez}, {Maury}, {Mauskopf}, {Mayet}, {Monfardini},
  {Motte}, {Pajot}, {Pascale}, {Peretto}, {Perotto}, {Pisano}, {Ponthieu},
  {Rev{\'e}ret}, {Rigby}, {Ritacco}, {Rodriguez}, {Romero}, {Roy}, {Ruppin},
  {Schuster}, {Sievers}, {Triqueneaux}, {Tucker}, \& {Zylka}}]{Bracco2017}
{Bracco}, A., {Palmeirim}, P., {Andr{\'e}}, P., {et~al.} 2017, \aap, 604, A52

\bibitem[{{Craig}(1979)}]{Craig1979}
{Craig}, I.~J.~D. 1979, \aap, 79, 121

\bibitem[{{Gladstone} {et~al.}(2016){Gladstone}, {Stern}, {Ennico}, {Olkin},
  {Weaver}, {Young}, {Summers}, {Strobel}, {Hinson}, {Kammer}, {Parker},
  {Steffl}, {Linscott}, {Parker}, {Cheng}, {Slater}, {Versteeg}, {Greathouse},
  {Retherford}, {Throop}, {Cunningham}, {Woods}, {Singer}, {Tsang},
  {Schindhelm}, {Lisse}, {Wong}, {Yung}, {Zhu}, {Curdt}, {Lavvas}, {Young},
  {Tyler}, {Bagenal}, {Grundy}, {McKinnon}, {Moore}, {Spencer}, {Andert},
  {Andrews}, {Banks}, {Bauer}, {Bauman}, {Barnouin}, {Bedini}, {Beisser},
  {Beyer}, {Bhaskaran}, {Binzel}, {Birath}, {Bird}, {Bogan}, {Bowman}, {Bray},
  {Brozovic}, {Bryan}, {Buckley}, {Buie}, {Buratti}, {Bushman}, {Calloway},
  {Carcich}, {Conard}, {Conrad}, {Cook}, {Cruikshank}, {Custodio}, {Ore},
  {Deboy}, {Dischner}, {Dumont}, {Earle}, {Elliott}, {Ercol}, {Ernst},
  {Finley}, {Flanigan}, {Fountain}, {Freeze}, {Green}, {Guo}, {Hahn},
  {Hamilton}, {Hamilton}, {Hanley}, {Harch}, {Hart}, {Hersman}, {Hill}, {Hill},
  {Holdridge}, {Horanyi}, {Howard}, {Howett}, {Jackman}, {Jacobson},
  {Jennings}, {Kang}, {Kaufmann}, {Kollmann}, {Krimigis}, {Kusnierkiewicz},
  {Lauer}, {Lee}, {Lindstrom}, {Lunsford}, {Mallder}, {Martin}, {McComas},
  {McNutt}, {Mehoke}, {Mehoke}, {Melin}, {Mutchler}, {Nelson}, {Nimmo},
  {Nunez}, {Ocampo}, {Owen}, {Paetzold}, {Page}, {Pelletier}, {Peterson},
  {Pinkine}, {Piquette}, {Porter}, {Protopapa}, {Redfern}, {Reitsema},
  {Reuter}, {Roberts}, {Robbins}, {Rogers}, {Rose}, {Runyon}, {Ryschkewitsch},
  {Schenk}, {Sepan}, {Showalter}, {Soluri}, {Stanbridge}, {Stryk}, {Szalay},
  {Tapley}, {Taylor}, {Taylor}, {Umurhan}, {Verbiscer}, {Versteeg}, {Vincent},
  {Webbert}, {Weidner}, {Weigle}, {White}, {Whittenburg}, {Williams},
  {Williams}, {Williams}, {Zangari}, \& {Zirnstein}}]{Gladstone2016}
{Gladstone}, G.~R., {Stern}, S.~A., {Ennico}, K., {et~al.} 2016, Science, 351,
  aad8866

\bibitem[{Glasser {et~al.}(1978)Glasser, Chapelle, \& Boettner}]{Glasser1978}
Glasser, J., Chapelle, J., \& Boettner, J.~C. 1978, Appl. Opt., 17, 3750

\bibitem[{{Hasenberger} {et~al.}(2018){Hasenberger}, {Lombardi}, {Alves},
  {Forbrich}, {Hacar}, \& {Lada}}]{Hasenberger2018}
{Hasenberger}, B., {Lombardi}, M., {Alves}, J., {et~al.} 2018, \aap, 620, A24

\bibitem[{{Hickstein} {et~al.}(2019){Hickstein}, {Gibson}, {Yurchak}, {Das}, \&
  {Ryazanov}}]{Hickstein2019}
{Hickstein}, D.~D., {Gibson}, S.~T., {Yurchak}, R., {Das}, D.~D., \&
  {Ryazanov}, M. 2019, arXiv e-prints [\eprint[arXiv]{1902.09007}]

\bibitem[{Hunter(2007)}]{matplotlib2007}
Hunter, J.~D. 2007, Computing In Science \& Engineering, 9, 90

\bibitem[{Jones {et~al.}(2001)Jones, Oliphant, Peterson, {et~al.}}]{scipy2001}
Jones, E., Oliphant, T., Peterson, P., {et~al.} 2001, {SciPy}: Open source
  scientific tools for {Python}, [Online; accessed July 9, 2019]

\bibitem[{{Kainulainen} {et~al.}(2014){Kainulainen}, {Federrath}, \&
  {Henning}}]{Kainulainen2014}
{Kainulainen}, J., {Federrath}, C., \& {Henning}, T. 2014, Science, 344, 183

\bibitem[{{Kr{\v{c}}o} \& {Goldsmith}(2016)}]{Krco2016}
{Kr{\v{c}}o}, M. \& {Goldsmith}, P.~F. 2016, \apj, 822, 10

\bibitem[{{Kursinski} {et~al.}(1997){Kursinski}, {Hajj}, {Schofield},
  {Linfield}, \& {Hardy}}]{Kursinski1997}
{Kursinski}, E.~R., {Hajj}, G.~A., {Schofield}, J.~T., {Linfield}, R.~P., \&
  {Hardy}, K.~R. 1997, \jgr, 102, 23429

\bibitem[{{Lada} {et~al.}(2017){Lada}, {Lewis}, {Lombardi}, \&
  {Alves}}]{Lada2017}
{Lada}, C.~J., {Lewis}, J.~A., {Lombardi}, M., \& {Alves}, J. 2017, \aap, 606,
  A100

\bibitem[{{Lee} {et~al.}(2015){Lee}, {Seon}, \& {Jo}}]{Lee2015}
{Lee}, D., {Seon}, K.-I., \& {Jo}, Y.-S. 2015, \apj, 806, 274

\bibitem[{{Li} \& {Burkert}(2016)}]{Li2016}
{Li}, G.-X. \& {Burkert}, A. 2016, \mnras, 461, 3027

\bibitem[{{Lombardi} {et~al.}(2014){Lombardi}, {Bouy}, {Alves}, \&
  {Lada}}]{Lombardi2014}
{Lombardi}, M., {Bouy}, H., {Alves}, J., \& {Lada}, C.~J. 2014, \aap, 566, A45

\bibitem[{{Minerbo} \& {Levy}(1969)}]{Minerbo1969}
{Minerbo}, G.~N. \& {Levy}, M.~E. 1969, Siam. J. Numer. Ann., 6, 598

\bibitem[{{Nielbock} {et~al.}(2012){Nielbock}, {Launhardt}, {Steinacker},
  {Stutz}, {Balog}, {Beuther}, {Bouwman}, {Henning}, {Hily-Blant}, \&
  {Kainulainen}}]{Nielbock2012}
{Nielbock}, M., {Launhardt}, R., {Steinacker}, J., {et~al.} 2012, \aap, 547,
  A11

\bibitem[{{Roy} {et~al.}(2014){Roy}, {Andr{\'e}}, {Palmeirim}, {Attard},
  {K{\"o}nyves}, {Schneider}, {Peretto}, {Men'shchikov}, {Ward-Thompson},
  {Kirk}, {Griffin}, {Marsh}, {Abergel}, {Arzoumanian}, {Benedettini}, {Hill},
  {Motte}, {Nguyen Luong}, {Pezzuto}, {Rivera-Ingraham}, {Roussel}, {Rygl},
  {Spinoglio}, {Stamatellos}, \& {White}}]{Roy2014}
{Roy}, A., {Andr{\'e}}, P., {Palmeirim}, P., {et~al.} 2014, \aap, 562, A138

\bibitem[{{Steinacker} {et~al.}(2016){Steinacker}, {Bacmann}, {Henning}, \&
  {Heigl}}]{Steinacker2016}
{Steinacker}, J., {Bacmann}, A., {Henning}, T., \& {Heigl}, S. 2016, \aap, 593,
  A6

\bibitem[{{Steinacker} {et~al.}(2005){Steinacker}, {Bacmann}, {Henning},
  {Klessen}, \& {Stickel}}]{Steinacker2005}
{Steinacker}, J., {Bacmann}, A., {Henning}, T., {Klessen}, R., \& {Stickel}, M.
  2005, \aap, 434, 167

\bibitem[{van~der Walt {et~al.}(2011)van~der Walt, Colbert, \&
  Varoquaux}]{numpy2011}
van~der Walt, S., Colbert, S.~C., \& Varoquaux, G. 2011, Computing in Science
  Engineering, 13, 22

\bibitem[{van~der Walt {et~al.}(2014)van~der Walt, Schönberger,
  Nunez-Iglesias, Boulogne, Warner, Yager, Gouillart, \& Yu}]{skimage2014}
van~der Walt, S., Schönberger, J.~L., Nunez-Iglesias, J., {et~al.} 2014,
  PeerJ, 2, e453

\bibitem[{{Zari} {et~al.}(2016){Zari}, {Lombardi}, {Alves}, {Lada}, \&
  {Bouy}}]{Zari2016}
{Zari}, E., {Lombardi}, M., {Alves}, J., {Lada}, C.~J., \& {Bouy}, H. 2016,
  \aap, 587, A106

\end{thebibliography}

\begin{appendix}
\section{Proof of generalised inverse Abel transform of a constant function}
\label{sec:proof}

The inverse Abel transform of a constant function can be generalised by restricting the requirement of spherical symmetry to only the z-axis. In the following, we show that as for the spherically symmetric inverse Abel transform, the column-density distribution $F(x,y)=c$ can be obtained by integrating the volume-density distribution $f(x,y,z)$ along the z-axis. A generalised form of Eq.~\ref{equ:iAt} can be written as
\begin{equation}
f(x,y,z) = \frac{c}{\pi} \frac{1}{\sqrt{R^2-x^2-y^2-z^2}}.
\end{equation}
Integrating along the z-axis within the appropriate limits and considering that $R$, $x$, and $y$ are constant along a given line of sight yields
\begin{equation}
\begin{aligned}
F(x, y)& = \int f(x, y, z) dz \\
&= \frac{c}{\pi} \int_{-\sqrt{R^2-x^2-y^2}}^{\sqrt{R^2-x^2-y^2}} \frac{1}{\sqrt{R^2-x^2-y^2-z^2}} dz \\
&= \frac{c}{\pi} \cdot \left.arcsin\left( \frac{z}{\sqrt{R^2-x^2-y^2}} \right) \right|_{-\sqrt{R^2-x^2-y^2}}^{\sqrt{R^2-x^2-y^2}} \\
&= \frac{c}{\pi} \cdot \pi = c.
\end{aligned}
\end{equation}

\section{Derivation of approximated density distribution}
\label{sec:derivations}

For voxels with $r>0$, we assume that the grid follows a spherical coordinate system and that for a given radius all voxels cover the same solid angle $\Omega_{voxel}=\Omega_{voxel}(r)=r^{-2}$.
\begin{equation}
\begin{aligned}
f_{voxel}(x,y,z) &= \int_{x_1}^{x_2} \int_{y_1}^{y_2} \int_{z_1}^{z_2} \frac{c}{\pi} \frac{1}{\sqrt{R^2-r^2}} dx dy dz\\
&\sim \int_{r_1}^{r_2} \frac{c}{\pi} \frac{1}{\sqrt{R^2-r^2}} r^2 \Omega_{voxel}(r) dr\\
&\sim \frac{c}{\pi} \left[\arcsin\left(\frac{r_2}{R}\right) - \arcsin\left(\frac{r_1}{R}\right)\right]
\end{aligned}
\end{equation}

For voxels at the centres of substructures, we start with the exact integral over three Euclidean dimensions. Each integration produces an inverse trigonometric function, which we approximate with its Taylor series up to first order:

\begin{equation}
\begin{aligned}
f_{voxel}^{central}(x,y,z) ={}& \int_{x_1}^{x_2} \int_{y_1}^{y_2} \frac{c}{\pi} \frac{1}{\sqrt{R^2-x^2-y^2}} \cdot \\ &\cdot \left.\arcsin\left(\frac{z}{\sqrt{R^2-x^2-y^2}}\right)\right|_{z_1}^{z_2} dx dy\\
\sim{}& \int_{x_1}^{x_2} \int_{y_1}^{y_2} \frac{c}{\pi} \frac{z_2-z_1}{R^2-x^2-y^2} dx dy\\
\sim{}& \int_{x_1}^{x_2} \frac{c}{\pi} (z_2-z_1) \frac{1}{R^2-x^2} \cdot \\ &\cdot \left.\mathrm{arctanh}\left(\frac{y}{\sqrt{R^2-x^2}} \right)\right|_{y_1}^{y_2} dx\\
\sim{}& \int_{x_1}^{x_2} \frac{c}{\pi} (z_2-z_1) \frac{(y_2-y_1)}{R^2-x^2} dx\\
\sim{}& \frac{c}{\pi} (z_2-z_1) (y_2-y_1) \frac{1}{R} \cdot \left.\mathrm{arctanh}\left(\frac{x}{R}\right)\right|_{x_1}^{x_2}\\
\sim{}& \frac{c}{\pi R}(z_2-z_2) (y_2-y_1) (x_2-x_1)
\end{aligned}
\end{equation}

\section{Reconstruction quality parameters}
\label{sec:parametertables}

\begin{table}[H]
\caption{Reconstruction quality parameters for spherically symmetric density distributions with a Gaussian profile}
\label{tab:sph_G}
\centering
\begin{tabular}{cccc}
\hline 
$\sigma$ & $f_{\Delta\rho/\rho_{in}}$ & $(\Delta\Sigma/\Sigma_{in})_{max}$ & $(\Delta\rho(r)/\rho_{in}(r))_{max}$ \\

[pixel] & [\%] & [\%] & [\%] \\ 
\hline 
3 & 71.2 & 0.16 & 5.9 \\ 
5 & 88.3 & 0.11 & 6.9 \\ 
8 & 96.1 & 0.16 & 3.2 \\ 
\hline 
\end{tabular} 
\end{table}

\begin{table}[H]
\caption{Reconstruction quality parameters for spherically symmetric density distributions with a Plummer profile}
\label{tab:sph_P}
\centering
\begin{tabular}{cccc}
\hline 
$R_P$ & $f_{\Delta\rho/\rho_{in}}$ & $(\Delta\Sigma/\Sigma_{in})_{max}$ & $(\Delta\rho(r)/\rho_{in}(r))_{max}$ \\

[pixel] & [\%] & [\%] & [\%] \\ 
\hline 
5 & 89.9 & 0.19 & 1.9 \\ 
10 & 97.6 & 0.11 & 2.1 \\ 
15 & 90.6 & 0.14 & 1.9 \\ 
\hline 
\end{tabular} 
\end{table}

\begin{table}[H]
\caption{Reconstruction quality parameters for spherically symmetric density distributions with a smooth-step profile}
\label{tab:sph_s}
\centering
\begin{tabular}{cccc}
\hline 
$a$ & $f_{\Delta\rho/\rho_{in}}$ & $(\Delta\Sigma/\Sigma_{in})_{max}$ & $(\Delta\rho(r)/\rho_{in}(r))_{max}$ \\

[pixel] & [\%] & [\%] & [\%] \\ 
\hline 
0.5 & 71.8 & 0.19 & 7.1 \\ 
1.5 & 83.5 & 0.14 & 5.0 \\ 
2.5 & 94.6 & 0.15 & 2.9 \\ 
\hline 
\end{tabular} 
\end{table}

\begin{table}[H]
\caption{Reconstruction quality parameters for prolate spheroidal density distributions with a Gaussian profile ($\sigma_x=8$)}
\label{tab:ell_G}
\centering
\begin{tabular}{cccc}
\hline 
$\sigma_x/\sigma_y$ & $f_{\Delta\rho/\rho_{in}}$ & $(\Delta\Sigma/\Sigma_{in})_{max}$ & $(\Delta\rho(r)/\rho_{in}(r))_{max}$ \\

  & [\%] & [\%] & [\%] \\ 
\hline 
1.5 & 79.1 & 0.08 & 10.5 \\ 
2.0 & 74.7 & 0.10 & 5.9 \\ 
2.5 & 73.6 & 0.12 & 3.2 \\ 
\hline 
\end{tabular} 
\end{table}

\begin{table}[H]
\caption{Reconstruction quality parameters for spherically symmetric density distributions with a Gaussian profile ($\sigma=8$) and a noise level $\sigma_{noise}=0.3$}
\label{tab:noise_0p3}
\centering
\begin{tabular}{cccc}
\hline 
min. step size & $f_{\Delta\rho/\rho_{in}}$ & $(\Delta\Sigma/\Sigma_{in})_{max}$ & $(\Delta\rho(r)/\rho_{in}(r))_{max}$ \\

  & [\%] & [\%] & [\%] \\ 
\hline 
0 & 45.9 & 0.27 & 22.3 \\ 
2$\sigma_{noise}$ & 48.4 & 0.50 & 13.5 \\ 
\hline 
\end{tabular} 
\end{table}

\begin{table}[H]
\caption{Reconstruction quality parameters for spherically symmetric density distributions with a Gaussian profile ($\sigma=8$) and a noise level $\sigma_{noise}=0.5$}
\label{tab:noise_0p5}
\centering
\begin{tabular}{cccc}
\hline 
min. step size & $f_{\Delta\rho/\rho_{in}}$ & $(\Delta\Sigma/\Sigma_{in})_{max}$ & $(\Delta\rho(r)/\rho_{in}(r))_{max}$ \\

  & [\%] & [\%] & [\%] \\ 
\hline 
0 & 38.9 & 0.43 & 27.5 \\ 
2$\sigma_{noise}$ & 40.7 & 0.77 & 16.1 \\ 
\hline 
\end{tabular} 
\end{table}

\section{Additional figure}
\label{sec:addfigures}

\begin{figure}[ht]
\begin{center}
  \resizebox{\hsize}{!}
  {\includegraphics{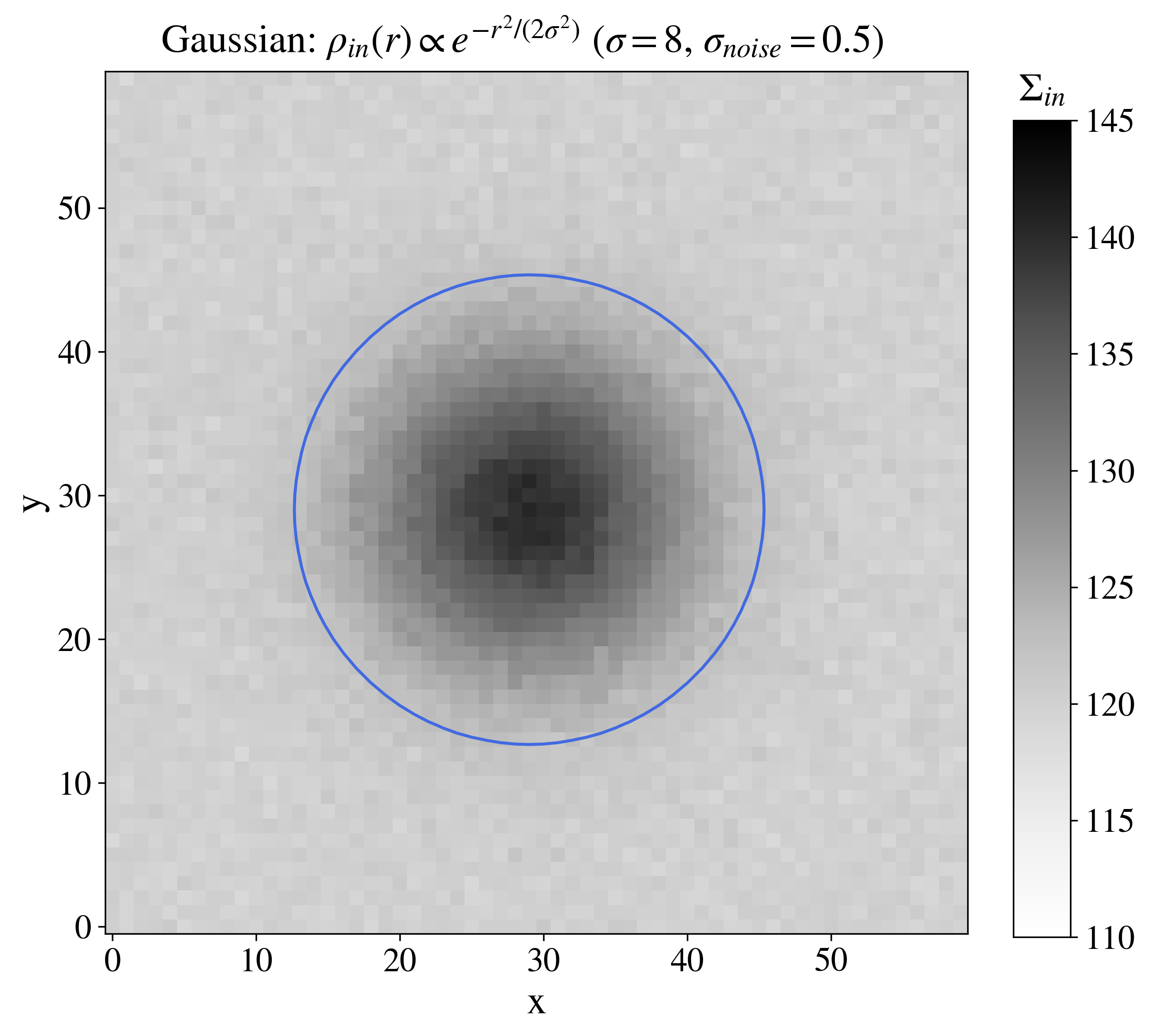}}
  \caption{Example of a column-density map of a spherical density distribution affected by noise. The blue circle indicates the radius at which the typical signal-to-noise ratio is 5.}
  \label{fig:val_noise_G8_ex}
\end{center}
\end{figure}

\end{appendix}

\end{document}